\documentclass[12pt,preprint]{aastex}
\usepackage{color}

\newcommand{\kmpers}{{{\rm \; km\;s}^{-1}}}

\newcommand{\magnitude}{{\rm \; mag}}

\newcommand{\eqw}{{W_{\lambda}}}

\received{}
\revised{}
\accepted{}

\shorttitle{Exoplanetary H-$\alpha$}

\shortauthors{Jensen et al.}

\begin{document}

\title{A DETECTION OF H$\alpha$ IN AN EXOPLANETARY EXOSPHERE}

\author{Adam G.~Jensen\altaffilmark{1}, Seth Redfield\altaffilmark{1}, Michael Endl\altaffilmark{2}, William D.~Cochran\altaffilmark{2}, Lars Koesterke\altaffilmark{2,3}, \& Travis Barman\altaffilmark{4}}

\altaffiltext{1}{Van Vleck Observatory, Astronomy Department, Wesleyan University, 96 Foss Hill Drive, Middletown, CT 06459; Adam.Jensen@gmail.com, sredfield@wesleyan.edu}
\altaffiltext{2}{University of Texas, Department of Astronomy, Austin, TX 78712; mike@astro.as.utexas.edu, wdc@astro.as.utexas.edu, lars@tacc.utexas.edu}
\altaffiltext{3}{Texas Advanced Computing Center, Research Office Complex 1.101, J.~J.~Pickle Research Campus, Building 196, 10100 Burnet Road (R8700), Austin, Texas 78758-4497}
\altaffiltext{4}{Lowell Observatory, 1400 West Mars Hill Road, Flagstaff, AZ 86001; barman@lowell.edu}

\begin{abstract}
We report on a search for H$\alpha$ absorption in four exoplanets.  Strong features at H$\alpha$ are observed in the transmission spectra of both HD 189733b and HD 209458b.  We attempt to characterize and remove the effects of stellar variability in HD 189733b, and along with an empirical Monte Carlo test the results imply a statistically significant transit-dependent feature of $(-8.72\pm1.48)\times10^{-4}$ integrated over a 16 \AA{} band relative to the adjacent continuum.  We interpret this as the first detection of this line in an exoplanetary atmosphere.  A previous detection of Ly$\alpha$ in HD 189733b's atmosphere allows us to calculate an excitation temperature for hydrogen, $T_{exc}=2.6\times10^4$ K.  This calculation depends significantly on certain simplifying assumptions.  We explore these assumptions and argue that $T_{exc}$ is very likely much greater than the radiative equilibrium temperature (the temperature the planet is assumed to be at based on stellar radiation and the planetary distance) of HD 189733b.  A large $T_{exc}$ implies a very low density that is not in thermodynamic equilibrium the planet's lower atmosphere.  We argue that the $n=2$ hydrogen required to cause H$\alpha$ absorption in the atmosphere is created as a result of the greater UV flux at HD 189733b, which has the smallest orbit and most chromospherically active central star in our sample.  Though the overall integration of HD 209458b's transmission spectrum over a wide band is consistent with zero, it contains a dramatic, statistically significant feature in the transmission spectrum with reflectional symmetry.  We discuss possible physical processes that could cause this feature.  Our remaining two targets (HD 147506b and HD 149026b) do not show any clear features, so we place upper limits on their H$\alpha$ absorption levels.

\end{abstract}

\section{INTRODUCTION AND BACKGROUND}
\label{s:intro}
Extrasolar planets that are known to transit their central star present an important opportunity for astronomers to study their atmospheres.  Following the first detection of a transiting exoplanet \citep{Charbonneau2000, Henry2000, Mazeh2000}, the first atmosphere of an exoplanet was discovered by \citet{Charbonneau2002}, who found sodium absorption in the atmosphere of HD 209458b.  In the ensuing years there have been additional detections in exoplanetary atmospheres of sodium \citep{Redfield2008, Snellen2008, Wood2011}, potassium \citep{Colon2010, Sing2011}, and molecules including H$_2$O, CH$_4$, CO, and CO$_2$ \citep{Barman2007, Beaulieu2008, Beaulieu2010, Grillmair2008, Swain2008, Swain2009a, Swain2009b, Swain2010, Tinetti2007}.\footnote{Note, however, that there is significant dispute over certain Spitzer results, most notably the \citet{Tinetti2007} detection of H$_2$O in HD 209458b; e.g., see \citet{Ehrenreich2007} for an alternative analysis of the data.}  These detections are presumed to come from the bound atmosphere of the host planets.

All atmospheric detections can be thought of as wavelength-dependent variations in apparent planetary radius, observed using either narrow-band photometry or spectroscopy.  With information about the absorbing species, this can be translated into a height for the absorbing material, which may indicate that the absorbing material extends to the unbound portion of an exoplanet's atmosphere, i.e., the exosphere.  This has been observed in multiple exoplanets, with HD 209458b being the most well-studied.  \citet{VidalMadjar2003,VidalMadjar2004} found \ion{H}{1}, \ion{C}{2}, and \ion{O}{1} in HD 209458b, while \citet{Linsky2010} confirmed the presence of \ion{H}{1} and detected \ion{Si}{3}.  \citet{Schlawin2010} also observed \ion{Si}{4} in this target.  Other exospheric detections include \ion{H}{1} in HD 189733b by \citet{LecavelierDesEtangs2010} and \ion{Mg}{2} and possibly other metals in WASP-12b by \citet{Fossati2010}.

Of particular importance to our study in this paper are the detections of \ion{H}{1} made for HD 209458b and HD 189733b.  \citet{VidalMadjar2003} observed HD 209458b with Space Telescope Imaging Spectrograph (STIS) onboard the {\it Hubble Space Telescope} ({\it HST}) and found a Ly$\alpha$ absorption level of $15\pm4$\%.  They calculated the Roche lobe of the planet to be at 2.7 Jupiter radii, but the height of the absorbing hydrogen required to produce 15\% absorption to be 4.3 Jupiter radii.  This indicates that the hydrogen overfills the Roche lobe and thus is escaping.  \citet{VidalMadjar2003} also calculated a simple escape model and placed a lower limit on the escaping material at $10^{10} {\rm \; g \; s^{-1}}$; the rate is higher if the absorption is saturated.

Additional observations of HD 209458b were presented in \citet{VidalMadjar2004}.  They found absorption in \ion{H}{1} of $5\pm2$\%, \ion{O}{1} of $13\pm4.5$\%, and \ion{C}{2} $7.5\pm3.5$\%.  The $15\pm4$\% \ion{H}{1} absorption measurement of \citet{VidalMadjar2003} is at higher resolution and detected at significant levels over a limited velocity range; \citet{VidalMadjar2004} estimate that this translates to $5.7\pm1.5$\% over the entire Ly$\alpha$ line at the lower resolution of the latter observations, and thus the two measurements are consistent.  The \ion{O}{1} and \ion{C}{2} transitions are multiplets including excited states that are blended in the lower-resolution data.  The higher-resolution data of \citet{VidalMadjar2003} show that the observed emission from the ground state \ion{O}{1} line is much weaker than the observed emission from \ion{O}{1}* or \ion{O}{1}** due to some combination of the inherent stellar emission and interstellar absorption.  This argues that the blended emission line in \citet{VidalMadjar2004} is dominated by \ion{O}{1}* and \ion{O}{1}** emission, and the 13\% absorption from the planetary atmosphere must primarily come from the excited transitions.  Similar arguments leave the carbon absorption ambiguous between ground-state \ion{C}{2} and \ion{C}{2}*.  Both results imply minimum densities in the atmosphere, because these excited states are collisionally populated.  \citet{VidalMadjar2003} argue that hydrodynamical escape or ``blow-off" is the likely mechanism, in which abundant \ion{H}{1} escapes at a high rate and other species (in this case the \ion{O}{1} and \ion{C}{2} are dragged hydrodynamically to fill the Roche lobe).  However, note that \citet{Ekenback2010} offer an alternative interpretation that involves an inflated hydrogen atmosphere and energetic neutrals (created through charge-exchange reactions between the stellar wind and the planet's atmosphere), but not evaporation.

\citet{LecavelierDesEtangs2010} used the spectroscopic mode of the Advanced Camera for Surveys (ACS) onboard {\it HST} to observe Ly$\alpha$ absorption in HD 189733b.  They measured a Ly$\alpha$ absorption level of $5.05\pm0.75$\% for in this target; the broadband transit depth is 2.4\%, meaning that the differential absorption level is 2.65\%.  \citet{LecavelierDesEtangs2010} argue that the \ion{H}{1} they detect must be unbound, and calculate constraints on the escape rate and EUV flux.  They find a range of $10^{9} {\rm \; g \; s^{-1}}$ to $10^{11} {\rm \; g \; s^{-1}}$ for the escape rate and 10 to 40 times the solar ionizing flux, with the best fit at $\mathrm{d}M/\mathrm{d}t = 10^{10} {\rm \; g \; s^{-1}}$ and $F(EUV) = 20 F_{\odot}(EUV)$.

It is important to note that all published cases of atmospheric detections in exoplanets are based on a single absorption line (or close doublet) or a single photometric detection (with a specific species only {\it inferred} as responsible) rather than a suite of lines, with one partial exception:   using low-resolution ACS spectra, \citet{Ballester2007} detected absorption in HD 209458b corresponding to the Balmer edge of \ion{H}{1}.\footnote{By the Balmer edge of \ion{H}{1} we mean the combination of the $n=2\rightarrow \infty$ transition edge at 3646 \AA{}, the associated bound-free absorption from the $n=2$ state at nearby but shorter wavelengths, and bound-bound absorption from $n=2 \rightarrow n \gg 2$ at nearby but longer wavelengths.}  \citet{Ballester2007} found $\sim$0.03\% absorption in addition to the planetary disk's baseline absorption.  Note, however, that this is a blended, conglomerate feature (especially when observed in low-resolution spectra) and not a single, resolved line.

Atomic hydrogen has been detected through its Ly$\alpha$ transition in the aforementioned targets and in the Balmer edge of HD 209458b, but not in the second-most prominent hydrogen line, H$\alpha$.  While the detection of a single line such as Ly$\alpha$ is important in its own right, it is very limited because it only reveals the presence of that atomic or molecular species and an extremely crude measure of density.  However, the detection of an additional line would provide interesting physical constraints on the density and temperature.  For example, measuring both Ly$\alpha$ and H$\alpha$ absorption (and assuming an absorption-density relationship to derive the density) would enable the calculation of an excitation temperature, $T_{exc}$.  We note that $T_{exc}$ does not necessarily equate to the local gas temperature.  However, it is still an interesting diagnostic of the physical conditions of an exosphere because a difference between the local gas temperature and $T_{exc}$ implies a deviation from LTE.

\citet{Winn2004} searched for H$\alpha$ in HD 209458b with the Subaru High Dispersion Spectrograph.  They were unable to detect any H$\alpha$ absorption, but placed an upper limit of 0.1\% over a 5.1 \AA{} band (to match the range over which Ly$\alpha$ is detected in HD 209458b).  In turn, \citet{Winn2004} placed a limit on the column density of hydrogen in the first excited state ($n=2$) of $N_2<10^{9} {\rm \; cm}^{-2}$ and an upper limit of $T_{exc}<8000$ K.  In their subsequent study, \citet{Ballester2007} argue that the Balmer edge detection in HD 209458b implies an absorption level of 0.3\% absorption in H$\alpha$, though not necessarily over the same waveband as measured by \citet{Winn2004}.

In this paper we report on our search for H$\alpha$ absorption in four exoplanetary systems previously examined for sodium and potassium absorption in \citet{Redfield2008} and \citet{Jensen2011}, hereafter R2008 and J2011, respectively.  Notably, we detect what appears to be H$\alpha$ absorption in the atmosphere of HD 189733b.  This is the first such detection of H$\alpha$ absorption in any exoplanetary atmosphere (or exosphere) and allows us to make a unique determination of $T_{exc}$ for the hydrogen associated with this planet.  We describe the observations and basic data reduction in \S\ref{s:obsdata}.  In \S\ref{s:analysis} we describe how we analyzed the reduced spectra including error determinations.  In \S\ref{s:results} we describe our basic results.  We discuss these results in \S\ref{s:discussion} and summarize in \S\ref{s:summary}.


\section{OBSERVATIONS AND DATA REDUCTION}
\label{s:obsdata}

\subsection{Description of Observations}
\label{ss:obsdescription}
We obtained observations of the planet-bearing systems HD 147506, HD 149026, HD 189733, and HD 209458 using the Hobby-Eberly Telescope (HET) from August 2006 to November 2008.  These data were used to make the first ground-based detection of an exoplanetary atmosphere by measuring Na I in HD 189733b (R2008).  These results were later confirmed in an semi-independent analysis of the same dataset (J2011), who also examined the spectra for K I absorption.  Recent HST observations have also confirmed the presence of \ion{Na}{1} in HD 189733b (Huitson et al., 2011, in preparation).

Details of the observations can be found in R2008 and J2011, but here we provide an overview.  Observations were made with the High-Resolution Spectrograph (HRS) at a nominal resolution of $R\sim60,000$.  Individual exposures were limited to 10 minutes in length in order to avoid saturation (and are in some cases shorter due to observing constraints).  Approximately 20\% of the exposures for each target were taken while its respective planet was in transit.  Specific details are provided in Tables \ref{table:observations} (observational statistics) and \ref{table:systemparams} (system and orbital parameters).

\begin{figure}[t!]
\begin{center}
\epsscale{1.00}
\plotone{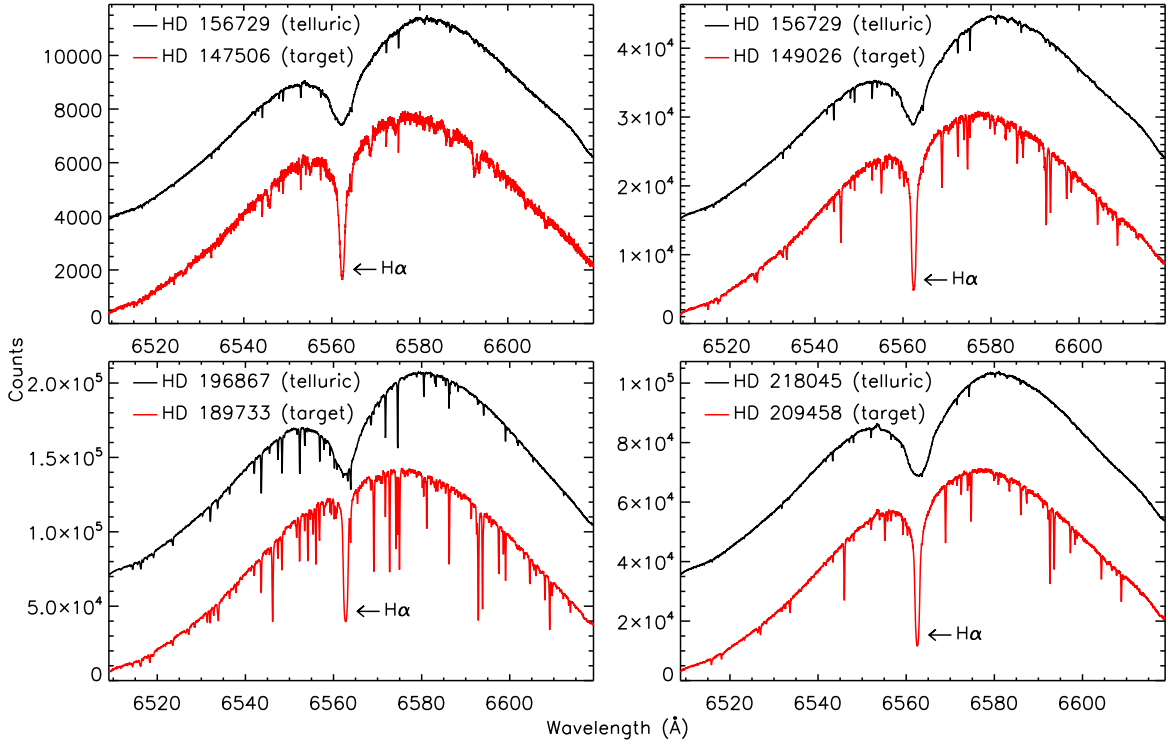}
\end{center}
\caption[]{Examples of the spectra in the echelle order that contains \ion{H}{1} $\lambda6562$ for all four targets, along with the spectra of their respective telluric standard stars from the same night of observation.  All spectra are reduced from the raw echelle images but are not processed with the additional steps we discuss in \S\ref{ss:reduction} (including removal of the H$\alpha$ feature from the telluric spectra, telluric correction from the primary target, and normalization).  The target spectra are shown on the count scale of the y-axis, but the telluric spectra are scaled to have the same maximum as the corresponding target spectra, and then arbitrarily shifted by 25\% of that same maximum.  The global shapes of the spectra are dominated by the blaze function of the echelle.}
\label{fig:haspecs}
\end{figure}

Table \ref{table:observations} also lists the telluric standard stars used for each observation.  Telluric standard observations were made either immediately before or after primary science observations, thus limiting spatial and temporal variations in air mass and water vapor levels.  Examples of the primary target and telluric standard spectra are shown in Fig.~\ref{fig:haspecs}.

\subsection{Data Reduction}
\label{ss:reduction}
The HRS instrument on the HET is an echelle spectrograph.  We used standard IRAF procedures to remove the bias level and scattered light, flatten the field, extract the apertures, and determine the wavelength calibration from the ThAr lamp exposures.  We extracted to one-dimensional spectra the primary science observations, the telluric standard star observations, and the ThAr exposures.

\begin{figure}[t!]
\begin{center}
\epsscale{1.00}
\plotone{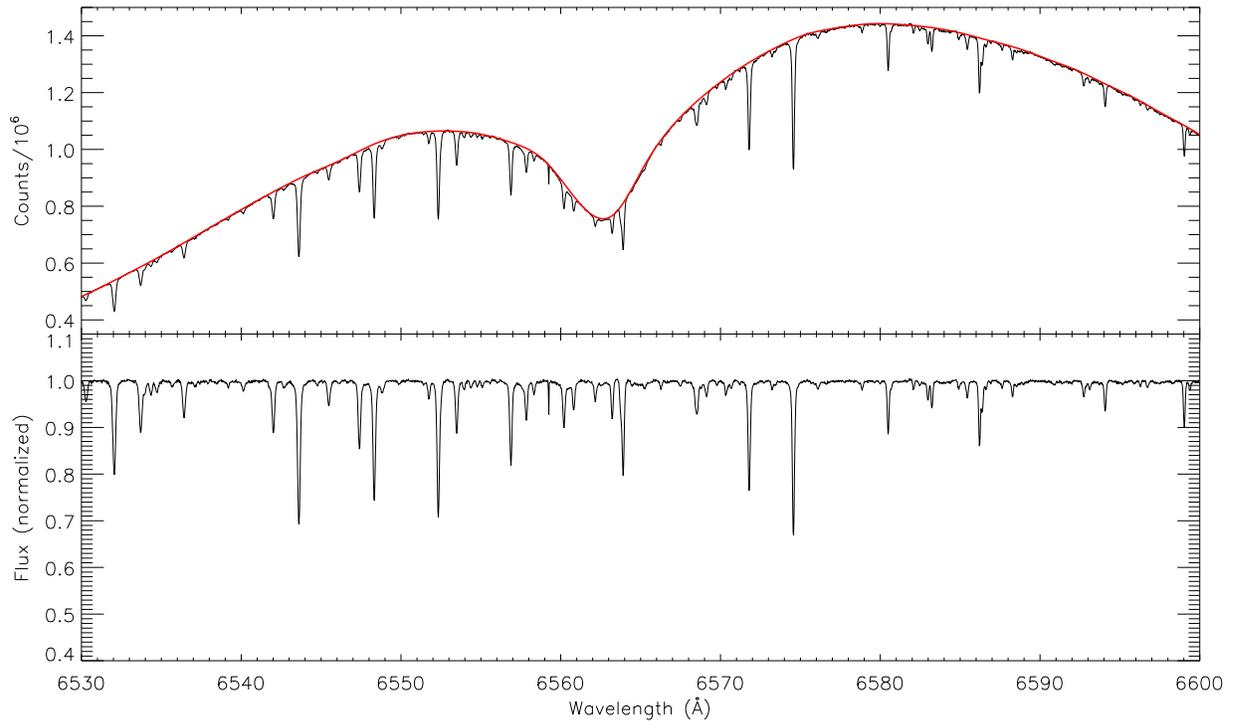}
\end{center}
\label{fig:splinefit}
\caption[]{A sample of a spline fit to one of the observations of the spectrum of star HD 196867, the telluric standard star for HD 189733.  The top panel shows the raw spectrum (scaled by $10^6$), along with its fit (thick red line), and the bottom panel shows the resulting residual, the normalized telluric absorption spectrum.}
\end{figure}

In order to perform a proper telluric correction, we must remove the imprint of the strong H$\alpha$ line in the telluric standard star.  We model a high-order spline (see Fig.~\ref{fig:splinefit}) through the entire spectral order containing the H$\alpha$ line in all of the telluric observations, then calculate an average model to be removed, via division, from these observations.  When dividing this model from each telluric observation, we also add a low-order polynomial to account for wavelength shifts and differing scales between observations.  After this division, we are left with a normalized telluric spectrum, free of the telluric standard's H$\alpha$ imprint.  The primary science observations are then corrected for telluric absorption using the normalized telluric spectra and standard IRAF procedures.  We note that we find no apparent telluric residuals in the resulting transmission spectra (see \S\ref{ss:transmission}).

We separate the exposures by whether they are in- or out-of-transit, and coadd spectra from a given group, leaving us with a master in-transit spectrum and a master out-of-transit spectrum.  Two additional issues arise in the coadding process.  First, by measuring unsaturated lines in the ThAr lamp exposures, we note that the resolution varies slightly as a function of time.  Secondly, we must apply small wavelength corrections, both an overall shift and a higher-order correction based on cross-correlating the stellar lines.  These two steps were described in more detail in J2011.  The final coadding is weighted by the errors at each point in the spectrum, with the errors based on photon statistics and read noise.

\section{ANALYSIS}
\label{s:analysis}

\subsection{Transmission Spectra}
\label{ss:transmission}
With our master in- and out-of-transit spectra we derive a transmission spectrum.  We perform a wavelength correction similar to that used in our coadding process (an overall shift plus a higher-order correction from cross-correlating stellar lines).  The spectrum is normalized, with the continuum defined several angstroms away from the line center.  We then calculate the transmission spectrum,
\begin{equation}\label{eq:st}
S_T=\left(\frac{F_{in}}{F_{out}}\right)-1 \;.
\end{equation}
We calculate a measurement of the absorption, $M_{abs}$, which we define as
\begin{equation}\label{eq:intsig}
M_{abs} = < \! \! S_T \! \! >_{c} - \frac{<\! \!S_T\! \!>_{b} + <\! \!S_T\! \!>_{r}}{2} ,
\end{equation}
where the $c$, $b$, and $r$ subscripts represent the central primary integration region, and the blue and red comparison integration regions, respectively.

\subsection{Estimating Errors with an Empirical Monte-Carlo Method}
\label{ss:montecarlo}
Because of the many steps in our reduction process, we suspect that systematic errors may be larger than the formally propagated statistical errors.  These systematic errors may come from either our reduction process or from astrophysical sources.  Systematic errors from our reduction process would include the telluric correction step, normalization of spectra, and our wavelength corrections.  The dominant astrophysical source of systematic error would be stellar variability.  To quantify the effect of these systematic errors, we use an ``Empirical Monte-Carlo" method (hereafter EMC).  The details of this method can be found in R2008 and J2011, but we summarize here.  We use our coadding process described in \S\ref{ss:reduction} on various subsets of the in- and out-of-transit data (always of the same type) and compare them to the master in- or out-of-transit spectra using our procedure for calculating transmission spectra.

The three important combinations we examine are:  (1) in-transit subsets vs.~the out-of-transit master spectrum or ``in-out"; (2) in-transit subsets vs.~the in-transit master spectrum or ``in-in"; and (3) out-of-transit subsets vs.~the out-of-transit master spectrum or ``out-out."  Each method will have some characteristic distribution, but we expect that the ``in-out" method will be centered around the master transmission spectrum measurement and the ``in-in" and ``out-out" methods will be centered around zero.  Additionally, we expect that the distribution of the ``out-out" method will provide the best estimate of the true systematic error (see R2008 and J2011 for more details).

\subsection{Confirming Detections with a Control Line}
\label{ss:controlline}
Another important check on our results is to see whether they are the systematic result of our coadding and transmission spectrum calculations near a strong stellar line.  To check this, we performed the above procedures on a strong stellar line of \ion{Ca}{1} at 6122 \AA{}, which we do not expect to manifest in exoplanetary atmospheres (calcium should condense out at the temperatures implied by models, e.g., into CaTiO$_3$).  If our method is robust even when performed on such a line, the results of the master measurement and the EMC distribution should be consistent with zero (see R2008 and J2011 for additional details).  These calculations were performed by J2011, and we do not repeat them here (see Fig.~2 and 5 of J2011).  In J2011, we did indeed find the master measurements and EMC distributions to be consistent with zero for all four systems, an indication that any absorption results on other lines are not simply an inevitable, or even likely, consequence of our analysis procedures.

\section{RESULTS}
\label{s:results}
\begin{figure}[t!]
\begin{center}
\epsscale{1.00}
\plotone{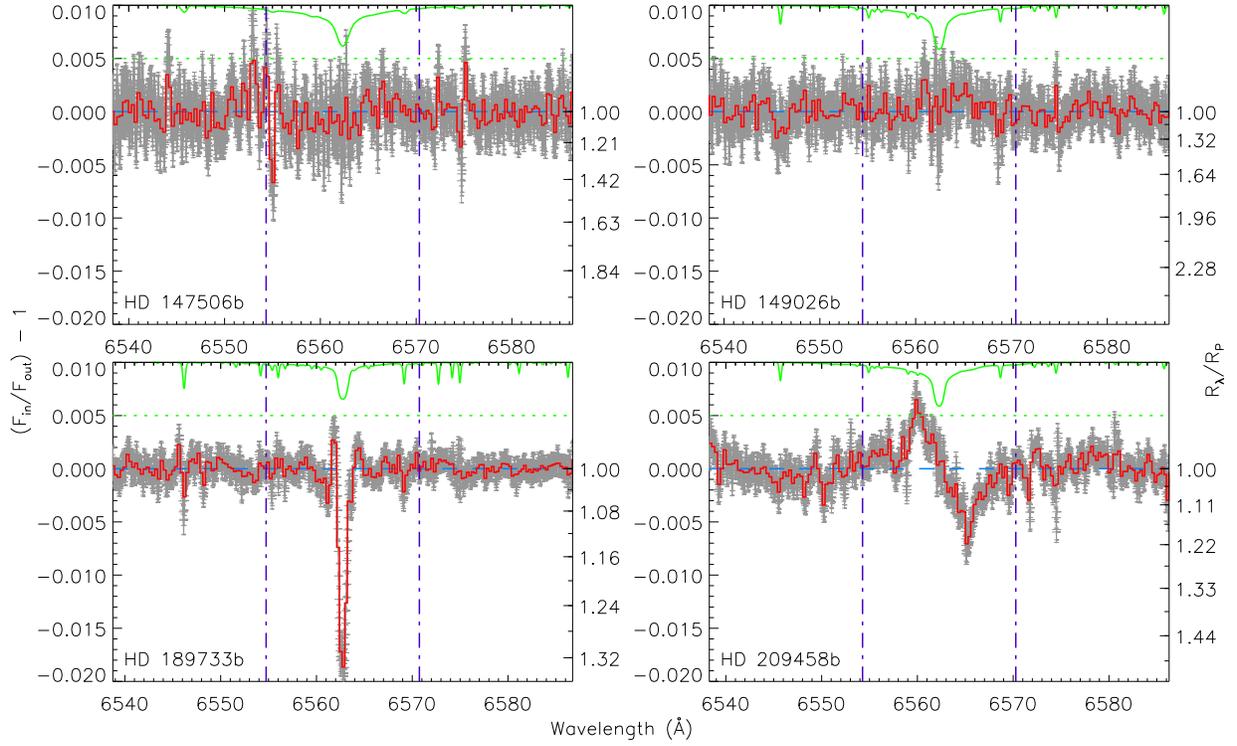}
\end{center}
\caption[]{The transmission spectrum.  The binned transmission spectrum is shown in red, with error bars (errors shown are the errors in the mean of the binned points).  For reference, the master out-of-transit spectrum of the star is shown at the top green, scaled down to fit the plot, with a dotted green line showing the zero level and the top of each plot being unity in the normalized spectrum.  A blue dashed line shows the zero point for the transmission spectrum.  The purple vertical dot-dashed lines define the bandpass that is integrated to make our absorption measurements; from these lines to the edges of the plot define the comparison bandpasses.}
\label{fig:hadiffspec}
\end{figure}

\begin{figure}[t!]
\begin{center}
\epsscale{1.00}
\plotone{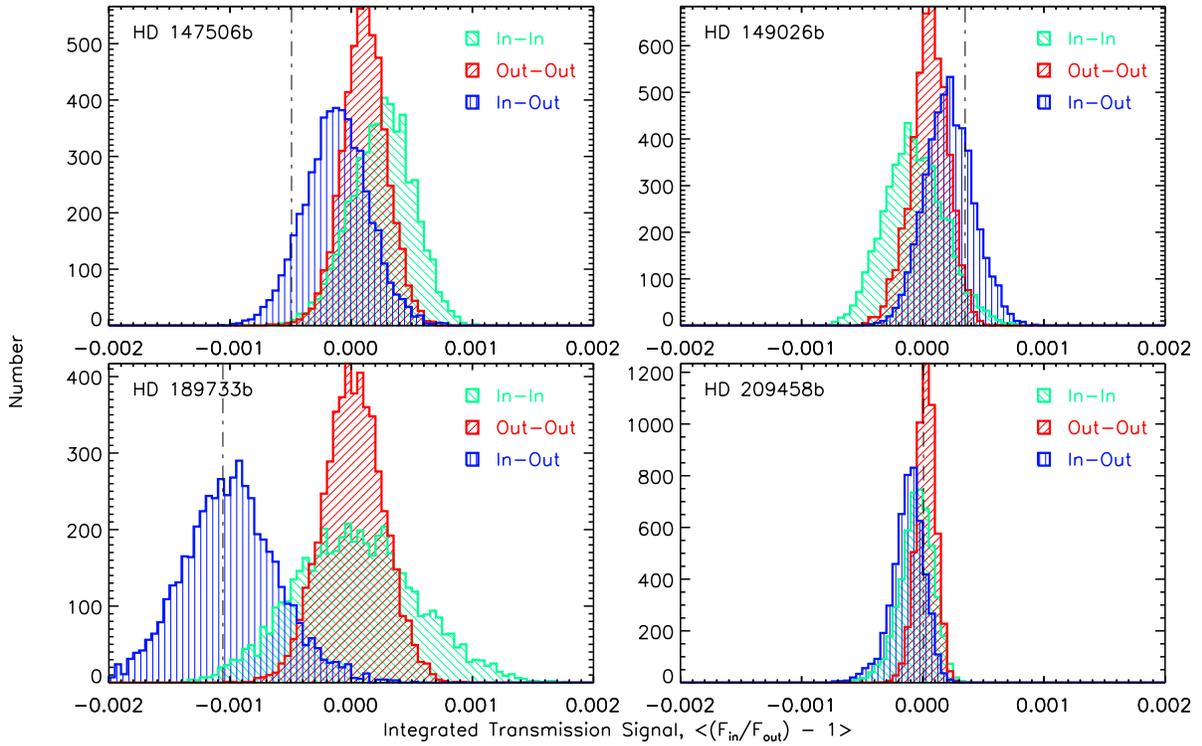}
\end{center}
\caption[]{Emprical Monte Carlot (EMC) analysis results.  The distributions for the three versions of the EMC (see \S\ref{ss:montecarlo}) are plotted in different colors, shown in the key:  green is the ``in-in" distribution, red is the ``out-out" distribution, and blue is the ``in-out" distribution.  5000 iterations are used in each case.  The master ``in-out" measurement is shown as a vertical dot-dashed line.}
\label{fig:haemc}
\end{figure}

The resulting transmission spectra from the reduction and analysis processes described above are shown in Fig.~\ref{fig:hadiffspec}.  The corresponding EMC analyses are shown in Fig.~\ref{fig:haemc}, and include ``in-in," ``in-out," and ``out-out" results.  In both figures, each panel corresponds to one of our targets.  For the commensurate results of the control line analysis of \ion{Ca}{1} described in \S\ref{ss:controlline}, see Fig.~2 and 5, respectively, of J2011.

Initially, we performed our integration (i.e., the calculation of $M_{abs}$ from Eq.~\ref{eq:intsig}) over a 16 \AA{} band.  We anticipate that H$\alpha$ may be broad, and therefore chose this integration band to be broader than the 12 \AA{} band used to measure \ion{Na}{1} in R2008, J2011, and \citet{Charbonneau2002} and the 8 \AA{} bands used to measure \ion{Ca}{1} and \ion{K}{1} in J2011.  However, because we compare this measurement to adjacent bands of equal width, a 16 \AA{} integration band requires a continuous spectrum of 48 \AA{} centered on the line.  The spectral width of the echelle orders is $\sim$100 \AA{}, and our selection of integration bandwidths are limited by this width and how close the line is to the center of the order.  However, we also define different bandpasses and perform the corresponding measurements for both the master measurement and the measurements of the EMC calculations.  In Table \ref{table:results} we show all the resulting $M_{abs}$ for a 16 \AA{} bandpass, which is the bandpass that is represented in the master transmission spectra and EMC analyses that are shown in Fig.~\ref{fig:hadiffspec}--\ref{fig:haemc}.  For comparison, Table \ref{table:results} also shows the results of a 6 \AA{} bandpass, which is discussed in \S\ref{ss:147-149results}.  A few additional custom bandpasses are defined to examine the unique feature in HD 209458b, which we describe in \S\ref{ss:209results}.

We also discuss stellar variability as it pertains to the features noted in HD 189733b and HD 209458b.  Our observations cover a wide range in time and allow for period analysis at the level of individual observations.  We find significant stellar variability at H$\alpha$ in the case of the star HD 189733, and this variability must be understood in order to properly interpret the transmission spectrum of the planet.

\subsection{HD 147506b and HD 149026b}
\label{ss:147-149results}
The transmission spectra of both HD 147506b and HD 149026b are relatively featureless compared to the noise level.  Notably, the strongest ``noise" feature is at approximately 6553 \AA{} in HD 147506b, which, at our settings, corresponds to the known location of a large group of bad pixels (roughly a $3\times3$ square of pixels) on the CCD chip.  Standard IRAF procedures and the telluric correction step largely take out the effects of these bad pixels in most individual exposures, but inspection here of the final transmission spectra shows that some residuals remain that are only obvious at high precision.

Though the transmission spectra for these two targets do not show any clear features that appear to be absorption or emission, the master integration measurements and corresponding EMC results show a slightly more complicated situation.  The master integration measurement for HD 147506b results in an absorption level that is 2.8$\sigma$ significant relative to the width of the ``out-out" EMC, which we assume is the best estimation of the total systematic and statistical error (see \S\ref{ss:montecarlo} and similar discussions in R2008 and J2011).  Notably, this master measurement is not well-aligned with the ``in-out" distribution, which is an exception in the 16 transmission spectra calculated between this work and J2011.  The best-fit centroid of the ``in-out" distribution is only 0.7$\sigma$ away from zero, again using the Gaussian $\sigma$ from the ``out-out" distribution as the error.  The ``in-in" distribution is also not quite centered at zero (though it is only approximately 1.0$\sigma$ away from zero using its own Gaussian width).

There are a few things we can say about the results for HD 147506b.  First, based on the master transmission spectra, it appears to be the only one of the four targets where the aforementioned section of bad pixels, affecting data points near $\sim$6553 \AA{}, comes into play in the final transmission spectrum.  Second, as we noted in J2011, HD 147506 is the faintest star in our sample, with the resulting lowest S/N (comparable to observations of HD 149026 but much lower than HD 189733 and HD 209458).  It also has the broadest and shallowest stellar features of the four stars in our sample, due to its $v \sin{i}$ of $20.8\pm0.3\kmpers$, which is $\geq3\times$ the other three stars \citep{Pal2010, Butler2006}.  This complicates our process of cross-correlating stellar lines of different spectra in order to make additional corrections to the wavelength solution (note, for example, that there is also a small alignment artifact at $\sim$6575 \AA{} in this target).  Therefore, it would have been reasonable to identify this target $a$ $priori$ as the most qualitatively uncertain, a conclusion we also reached in J2011 where similar issues were seen in the transmission spectrum of this target.  Finally, we note that we obtain different result when we integrate over a narrower bandpass.  Our default bandpass is 16 \AA{}; we also calculate a 6 \AA{} bandpass, which is chosen so that the blue comparison band misses the $\sim$6553 \AA{} noise feature.  This results in the master transmission spectrum integration being reduced; note that this is an average value, so the two are already normalized despite the different bandpasses.  The distribution of the ``out-out" EMC is also wider, so the significance drops to 0.58$\sigma$.  The consistency of the master measurement and the ``in-out" distribution is slightly improved in an absolute sense, and dramatically improved relative to the now-broader widths of the both the ``in-out" and ``out-out" distributions.  Furthermore, note that in both cases, the $\sim$6553 \AA{} noise feature may affect the overall normalization of the transmission spectrum.

In the transmission spectrum HD 149026b there is a slight ``emission" bump just to the red of the stellar H$\alpha$ line center.  This turns out to be significant at the 2.0$\sigma$ level.  Unlike the unusual EMC distributions (in that there are apparent internal inconsistencies) of HD 147506b described above, in this case the ``in-in" and ``out-out" distributions are both consistent with zero, and the ``in-out" distribution is consistent with the master integration measurement.  Changing the bandpass also does not affect these results in the same way that we described above for HD 147506b.  The feature is visually subtle and only 2.0$\sigma$, so we are disinclined to associate this with a real feature.  However, if the feature is real, the primary way that an emission feature could be observed in a transmission spectrum is by induced stellar activity from a star-planet interaction (SPI), which might occur on the portion of the stellar surface facing the planet.  We discuss this more in \S\ref{sss:189description} and \S\ref{sss:209broadhashift}.

\subsection{HD 189733b}
\label{ss:189results}
We detect a dramatic absorption-like feature in the transmission spectrum of HD 189733b, with a value of $M_{abs}=(1.06\pm0.24)\times10^{-3}$ (integrated over 16 \AA{} and $M_{abs}$ defined as in Eq.~\ref{eq:intsig} and using the standard deviation of the ``out-out" EMC distribution as our error).  This detection is significant at a 4.4$\sigma$ level.  The EMC results place important constraints on any otherwise undetected systematic errors, including any remaining atmospheric residuals of our telluric correction and planet-independent stellar variability.  The statistical significance of this result argues strongly that this is a real, transit-dependent feature rather than a reduction or analysis artifact.  We also note that when we inspect the transmission spectra of the several thousand EMC iterations visually, the results are qualitatively consistent with the measurements:  all ``in-out" iterations consistently show an absorption feature in transmission, whereas the ``in-in" and ``out-out" iterations show variation in the residuals of emission or absorption in the H$\alpha$ line, but the variation is centered about zero and smaller in magnitude than average magnitude of the ``in-out" absorption.  The variation in the ``in-in" and ``out-out" iterations is probably due to stellar variability, which we discuss below.  Another qualitative argument in favor of the HD 189733b result being a real phenomenon associated with transit is the null results in HD 147506b and HD 149026b.  Similar to the \ion{Ca}{1} control line in these targets, this shows that an artificial feature is not an inevitable result due to some effect of our transmission spectrum procedure performed on a strong stellar line.


However, the level of variability in the ``in-in" and ``out-out"---which we noted above was less than the magnitude of the absorption feature in the master transmission spectrum---is worth further inspection.  The EMC takes a random sample of the observations and is unlikely to sample all observations from a given night.  Therefore, if there is significant stellar variability on time scales greater than the length of the individual observations but shorter than the time spanned by different sets of observations (e.g., nights), then the EMC as initially constructed may not adequately characterize the variability.  Therefore, we make an additional attempt to do so by calculating a transmission spectrum and integrated absorption measurement for each individual out-of-transit observation relative to the master out-of-transit spectrum.  This assumes that the master out-of-transit spectrum sufficiently samples any variability and averages it out; using only out-of-transit observations explicitly removes the dependence of planetary transit from the calculation.  This provides us with a time series representation of the variability, on which we can perform period analysis.

\begin{figure}[t!]
\begin{center}
\epsscale{1.00}
\plotone{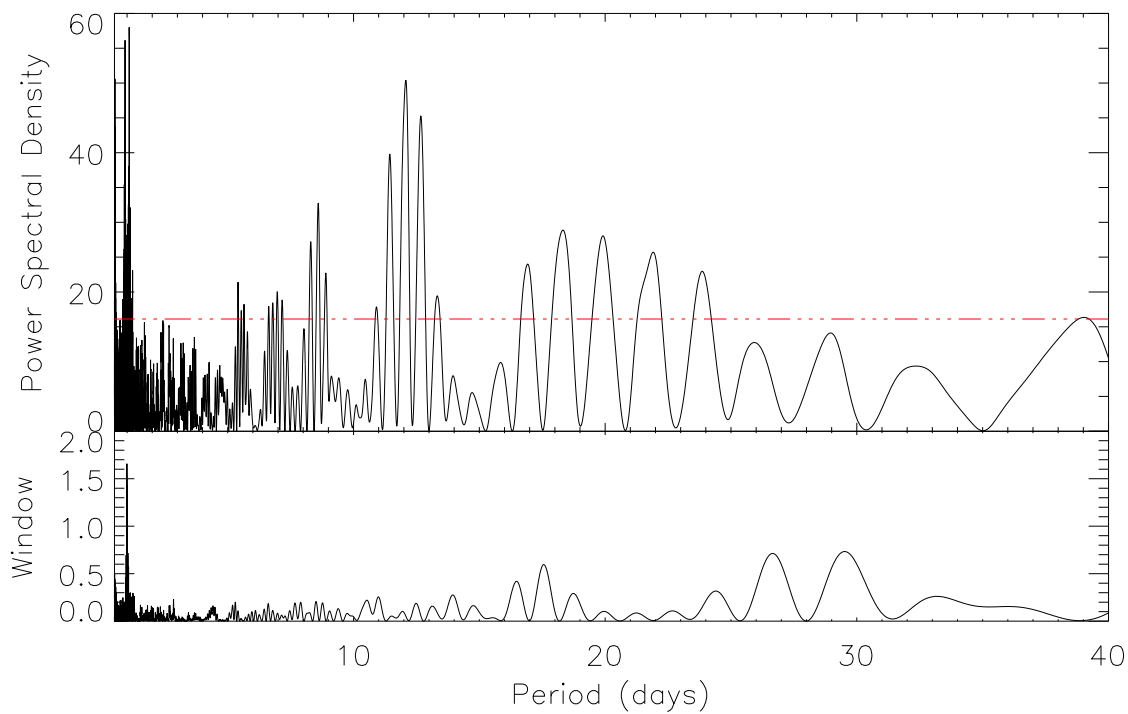}
\end{center}
\caption[]{The Lomb-Scargle periodogram of observations of HD 189733 H$\alpha$ individual ``transmission spectra" (i.e., individual out-of-transit spectra relative to the master out-of-transit).  The horizontal red dashed-dotted line is the 0.3\% ($\sim3\sigma$) false alarm probability level.  The dominant peaks are at $\sim1.086$ and $\sim12.077$ days, with many additional correlated peaks.  The window function is also shown, with a peak at 0.9967 days, which is expected because the timing of the observations will create a peak at the sidereal period of 0.9972 days.}
\label{fig:189periodogram}
\end{figure}

We used a Lomb-Scargle periodogram to analyze this time series, which is shown in Fig.~\ref{fig:189periodogram}, along with the window function associated with our data.  There are many periods with power spectral density (PSD) of greater than 3$\sigma$ significance, according to both analytic determination and white noise simulation estimations of the false alarm probability.  The strongest group of peaks in the periodogram, in terms of PSD, is at high frequency ($\sim$1 day); the second strongest set of peaks is near $\sim$12 days; the third strongest set of peaks is at $\sim$8.5 days; and more are found at one-half of the $\sim$12 day group, as well as lower frequencies of $\gtrsim$16 days.  In contrast, similar periodograms for HD 147506 and HD 149026 do not show any variability that exceeds the 3$\sigma$ false alarm threshold (we deal with HD 209458 in \S\ref{ss:209results}).  Given that HD 189733 is the most chromospherically active star in our small sample (see discussion in \S\ref{sss:189interpretation}), this might be expected.  Of the periods just described, the $\sim$12 day period is the one with a clear physical interpretation.  \citet{Bouchy2005} estimated a rotational period of $\sim$11 days, based on the $v \sin{i}$ and the assumption that we are viewing the star along its plane of rotation.  Similarly, \citet{HebrardLDE2006} found microvariability at several periods, with the dominant period to be at 11.8 days, which they interpreted as the rotation period.  ``Quasi-periodic" flux variations were measured by \citet{Winn2007rot189} with a period of 13.4 days.  \citet{Fares2010} suggest the discrepancy between these latter two results may be due to the temporal variation in stellar spots coupled to differential rotation, and find an equatorial period of $11.94\pm0.16$ days.

\citet{Fares2010} observed \ion{Ca}{2} and H$\alpha$ and found significant variability.  For each line, they observe over three epochs and calculate a best-fit period.  The H$\alpha$ best-fit periods are $11.6^{+0.3}_{-0.2}$, $14.4^{+1.5}_{-1.3}$, and $13.8^{2.2}_{-2.3}$ days for observations in 2006 June and August, 2007 June, and 2008 July (though the authors note that the phase coverage is poor for the 2006 data).  Note that all three best-fit H$\alpha$ periods are consistent with our period of 12.09 days (the period with the largest PSD between 10 and 15 days) to within 2$\sigma$; 12.077 days is also consistent to within $1\sigma$ with their derived equatorial period.  

Notably, there is also a straightforward explanation for the high-frequency peaks.  The window function peaks at 0.9967 days, which is expected because the timing of the HET observations is correlated with the length of the sidereal day, which is 0.9972 civil days.  The two peaks in the periodogram that are closest to this peak in the window function are also the two most powerful peaks in the periodogram, at 1.086 and 0.920 days.  These peaks are, to high precision, plus and minus one 12.077 day period in frequency space away from the window function peak.  This strongly suggests that these peaks are due to aliasing of the longer period in conjunction with the window function peak of 1 sidereal day, and therefore the power in these peaks should be disregarded.

\begin{figure}[t!]
\begin{center}
\epsscale{1.00}
\plotone{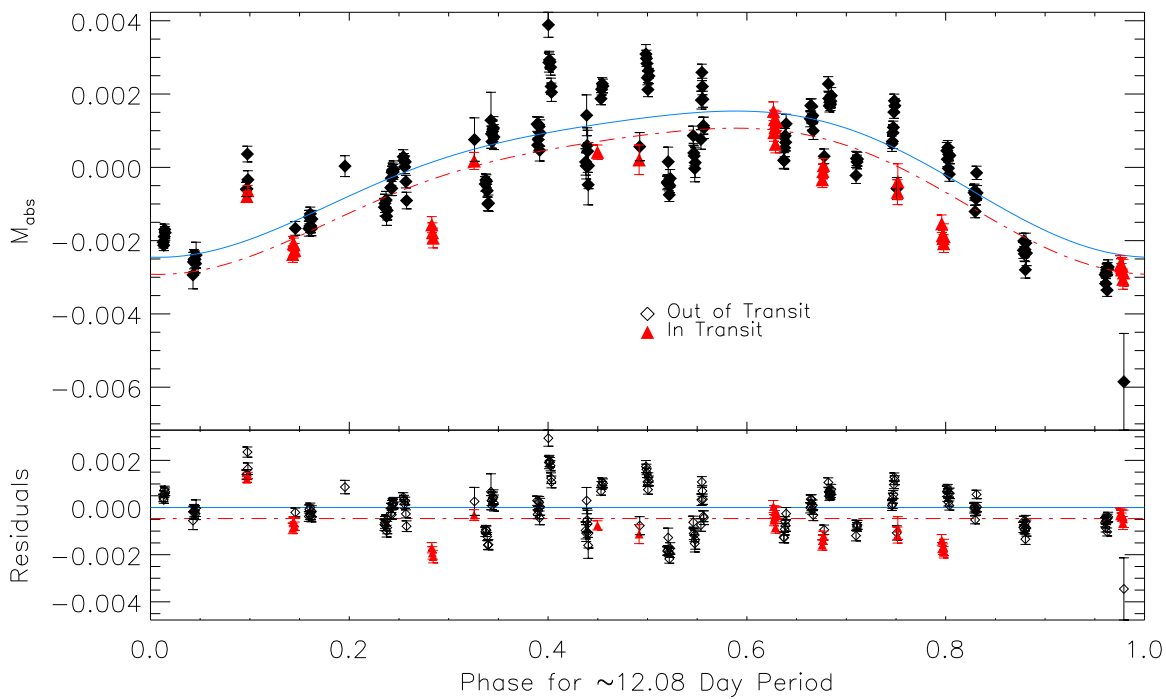}
\end{center}
\caption[]{The individual out-of-transit transmission spectra of HD 189733 phased to the $\sim12.077$ day period, which corresponds closely to the stellar rotation period.  The blue solid line is a low-order Fourier series fit to the out-of-transit points.  The red dashed-dotted line is that same out-of-transit fit where we introduce an offset to fit the in-transit points.  The systematic excess absorption seen in the in-transit data relative to the out-of-transit observations is evident.}
\label{fig:189sinefits}
\end{figure}

\begin{figure}[t!]
\begin{center}
\epsscale{1.00}
\plotone{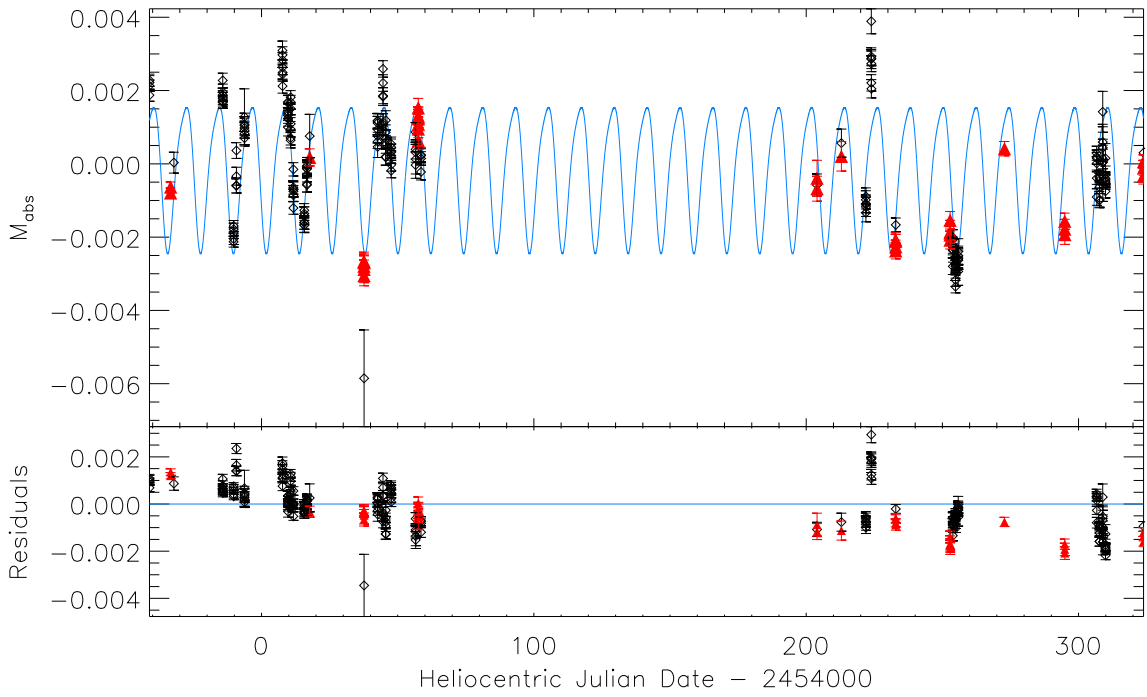}
\end{center}
\caption[]{Same as Fig.~\ref{fig:189sinefits} but as a function of absolute time instead of phase.  (No fit to the in-transit points is shown.)}
\label{fig:189sinefitstime}
\end{figure}

We attempt to remove the variability from our master transmission measurement by making a fit based on only the out-of-transit variability.  Then we can estimate an additive calibration term for any observation (in- or out-of-transit) by phasing it to this function.  These calibration terms are then averaged for all in-transit observations in order to calculate a calibrated master transmission spectrum integration, and also for each iteration of the EMC.  We show a fit to data phased to the 12.077 day period in Fig.~\ref{fig:189sinefits}, using a low-order Fourier series.  In Fig.~\ref{fig:189sinefitstime} we show this same fit as a function of absolute time.  Both Fig.~\ref{fig:189sinefits} and \ref{fig:189sinefitstime} show the general tendency of the in-transit residuals to be below the out-of-transit residuals.  When we calculate a new periodogram with the residuals, the PSD of the $\sim1$ day high-frequency peaks is greatly reduced---the two highest-power peaks disappear and the highest remaining $\sim1$ day period with the largest PSD is at the peak of the window function.  The other significant peaks of higher frequency than the 12.077 day period that we noted above are also significantly reduced.  We limit the fit to two terms in the Fourier series (i.e., the function is a linear combination of sines and cosines with periods of $P$ and $P/2$ where $P$ is the 12.077 day period) based on what we see in the periodogram; if more terms are allowed, the function begins to fit for some of the periodicity caused by our sampling frequency as characterized by the window function.


The calibrated EMC and master measurement gives us a new result of $(8.72\pm1.48)\times10^{-4}$.  The error from the EMC in this case has gone down significantly because we are explicitly fitting for the dominant source of astrophysical noise.  HD 189733 is the brightest star in our sample with the greatest number of observations and therefore should have the highest S/N and lowest corresponding error if astrophysical noise is removed.  Prior to calibrating the error, it was largest for HD 189733b.  The calibrated error is smaller than for HD 147506b and HD 149026b with their fainter central stars, but larger than for HD 209458b and its comparably bright star.  Using the uncalibrated measurements, our result was significant at a $4.4\sigma$ level; the calibrated result is significant to $5.9\sigma$, and the EMC results behave as we expect---the widths of the three EMC distributions are each reduced by a reasonably consistent fraction of between 39 and 46\%, the master ``in-out" measurement is consistent with the ``in-out" EMC, and the ``in-in" and ``'out-out" EMCs are consistent with zero.


\subsection{HD 209458b}
\label{ss:209results}
HD 209458 is equally intriguing for different reasons.  We do not detect an overall absorption or ``emission" signal, as the master integration measurement over the 16 \AA{} band is consistent with zero.  All three variations of the EMCs are centered on zero, although the ``in-in" and ``in-out" distributions are slightly asymmetric.  However, the transmission spectrum shows a dramatic feature with a ``spike" to the blue of stellar line center and a ``dip" to the red.  Both features peak at roughly 0.5\% absolute deviation from zero, and are several angstroms wide.  The feature is roughly symmetric if a reflection is applied about the zero point and line center.  When we use our EMC process and shift the integration bands to correspond to these features, the EMCs confirm that these are consistent features that are highly significant.  In order to explore these features which appear to reflect about the stellar H$\alpha$ line center, we examined $<\! \!S_T\! \!>$ in two 5 \AA{} bands with their edges centered on the stellar line center (unlike the 16 \AA{} and 6 \AA{} bandpasses of Table \ref{table:results} which are {\it centered} on the stellar line); 5 \AA{} is the approximate width of the spike and dip.  We find that in the blueward spike, $<\! \!S_T\! \!>$ (omitting the comparison bands) is an 6.6$\sigma$ ``emission" feature, and in the redward dip, $<\! \!S_T\! \!>$ is a 6.1$\sigma$ absorption feature.  The corresponding ``in-in" and ``out-out" distributions are consistent with zero to within better than $0.5$$\sigma$.  Visual inspection of the EMC iterations qualitatively confirms these results---i.e., these features are consistently seen in the various ``in-out" iterations and not present in the ``in-in" and ``out-out" iterations.  We also note that if we integrate over the range of $-130$ to $+100\kmpers$---the range over which \citet{VidalMadjar2003} detected significant Ly$\alpha$ absorption, and the range over which \citet{Winn2004} search for H$\alpha$ absorption---we arrive at a value for $<\! \!S_T\! \!>$ of $1.27\times10^{-3}$.  The positive value is due to the asymmetry of the velocity limits favoring the blueward spike; over a similar range, \citet{Winn2004} placed an upper limit of 0.1\%.  We suggest an explanation for the discrepancy later in this section.

It is difficult to explain apparent ``emission" features in a transmission spectrum.  However, it is not impossible for such features to occur.  Because of the nature of normalizing $S_T$ to 0, an ``emission" feature indicates a wavelength-dependent apparent planetary radius ($R_{\lambda}/R_P$) of less than one.  However, this assumes a constant stellar disk intensity.  If the stellar disk intensity changes as a function of both wavelength and time, wavelength-dependent features may be observed in the cumulative transmission spectrum.  For example, in the Rossiter-McLaughlin effect, the exoplanetary disk blocks different portions of the stellar disk, and each portion of block stellar surface has a different radial velocity relative to the observer due to stellar rotation.  This creates wavelength-dependent variations in the disk intensity.  The apparent symmetry of the HD 209458b feature also indicates the possibility that it is due to some sort of velocity misalignment between two absorption features that is not fully corrected in our transmission spectrum calculation (cf.,~the artifact at $\sim$6575 \AA{} in the transmission spectrum of HD 147506b).  In the case of HD 209458b, this would have be a very broad feature that is misaligned between the in- and out-of-transit spectra.  The misalignment would also need to be relative to the core of the stellar H$\alpha$ line, as there is no residual indicating that the strong central H$\alpha$ cores are misaligned in our spectra.  While we cannot completely rule out a reduction artifact, if such a misalignment is responsible for this feature, its consistency in the various ``in-out" EMC iterations argues for a genuine physical origin.  This could perhaps be due to SPI-related stellar variability.  Another hypothesis is that the ``broad" component of the stellar H$\alpha$ line actually includes absorption from excited hydrogen that has been lost by the planet, remains around the star in some sort of toroidal shape, and is influenced by the planet's orbit.  We discuss potential interpretations more in \S\ref{ss:HD209458b}.

\begin{figure}[t!]
\begin{center}
\epsscale{1.00}
\plotone{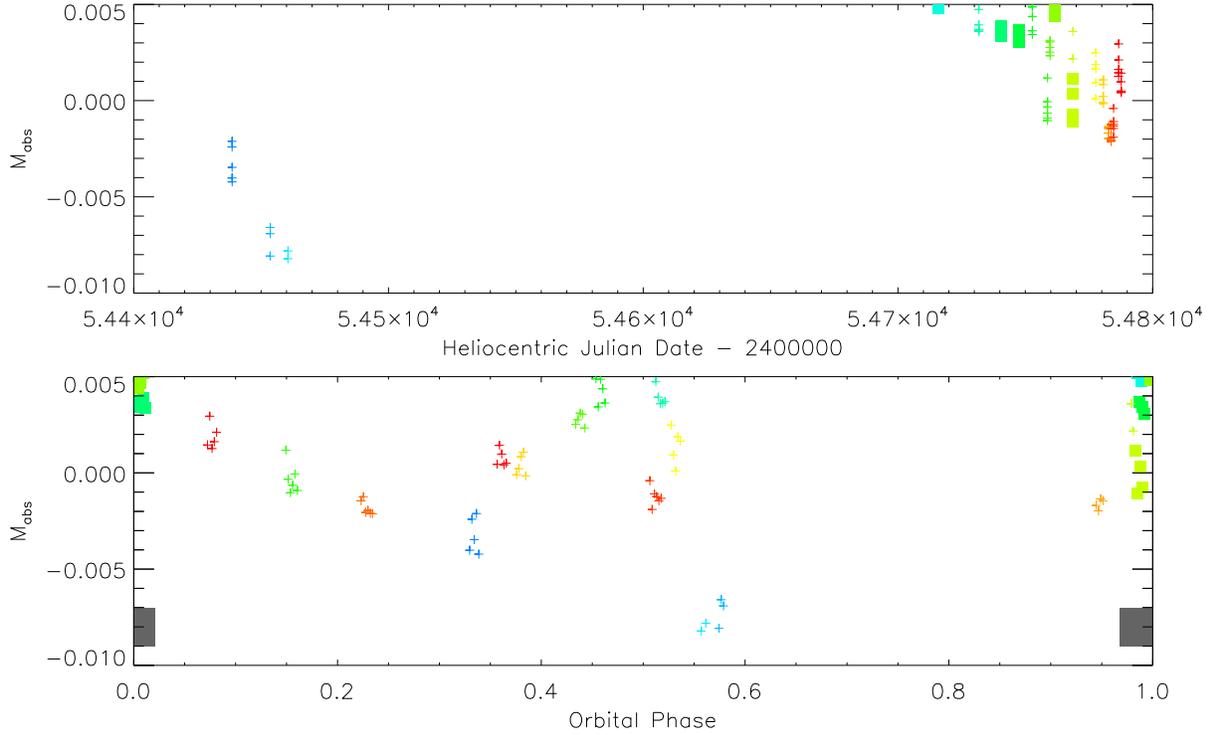}
\end{center}
\caption[]{This plot shows, for HD 209458, the individual transmission spectrum measurements.  The $M_{abs}$ here is for a 5 \AA{} band immediately to the blue of H$\alpha$ line center, quasi-normalized to 5 \AA{} bands on either side.  Therefore, a positive measurement is expected for what is seen in the master HD 209458b transmission spectrum, and a negative measurement for an inverted feature.  Crosses are out-of-transit observations relative to the master out-of-transit spectrum; squares are in-transit observations relative to the same master out-of-transit spectrum.  The top panel shows these measurements as a function of absolute time, while the bottom panel is phased to the orbital period of the planet.  The color-coding of the top panel is kept for the bottom panel; note that the possible phase correlation in the bottom panel does not appear to have any correlation to date.  The grey bar shows the phase range of the measurements by \citet{Winn2004}; the y-value of this bar is arbitrary.  If the feature in the master transit spectrum has a physical origin (rather than an as-yet undetermined data reduction explanation), and if the feature is phase-correlated but not strictly transit-correlated (as indicated by the bottom panel), the limited phase range of the observations of \citet{Winn2004} compared to ours explains why we see this feature and they do not.}
\label{fig:209phase}
\end{figure}

As with HD 189733b, we investigated whether or not there is variability beyond just transit that is not picked up by the EMCs.  This is represented in Fig.~\ref{fig:209phase}, using the 5 \AA{} integration band to the blue of the H$\alpha$ line as the measure of $M_{abs}$.  Positive measurements therefore mean that there is a blue spike and a red dip relative to H$\alpha$; negative measurements would indicate an inverted feature.  Notably, there does seem to be a phase correlation, as the blue spike is prominent during transit and near secondary eclipse (orbital phase $\phi\sim0.5$), but at intermediate values ($\phi\sim0.25$ and $\phi\sim0.75$), there is either no feature or an inverted feature (we examined the spectra and confirmed visually that these descriptions are the correct interpretation of $M_{abs}$, which again, in this case, refers to the offset measurement centered on the blue spike).  Such features apparently do not show up as readily in the ``in-in" and ``out-out" EMCs due to the fact that observations are selected individually and there are very few, if any, iterations that are dominated by a single night.  If the phase correlation seen in Fig.~\ref{fig:209phase} is real, and the effect seen in the master transmission spectrum is not strictly a transit effect but a broader phase effect, then it also explains the fact that \citet{Winn2004} did not note this feature, as they made observations over only $\sim5$ hours on one night, through a single transit, and did not have the wide range of phase coverage that we present here.





\section{DISCUSSION}
\label{s:discussion}

\subsection{Limits on H$\alpha$ in HD 147506b and HD 149026b}
\label{ss:147-149limits}
As described in \S\ref{ss:147-149results}, in neither HD 147506b nor HD 149026b do we detect clear H$\alpha$ absorption.  The atmospheres of these planets are not well-studied, with no Ly$\alpha$ detections (or null results),\footnote{There are observations of HD 149026b using ACS in the HST archives that cover Ly$\alpha$.  However, these observations were taken while STIS---which is, generally speaking, preferable to ACS for spectroscopy and therefore for detecting wavelength-dependent absorption in the UV---was not operating, and there are no published results associated with this program.} and comparatively little IR work.  No Spitzer observations have been taken of HD 147506b \citep{SeagerDeming2010}, and only 8 $\mu$m Spitzer observations have been taken of HD 149026b \citep{Harrington2007, Knutson2009}.  In an analysis of the potential connection between stellar activity and the presence or lack of atmospheric temperature inversions, \citet{Knutson2010} categorized these two planets' atmospheres as ``unknown" with respect to temperature inversions.

In J2011 we examined these spectra for absorption at \ion{Na}{1} $\lambda\lambda$5889,5895 and \ion{K}{1} $\lambda$7698.  We did not find clear evidence of absorption from either species, though we reported the suggestion of \ion{Na}{1} absorption in HD 149026b, significant at the 1.7$\sigma$ level, with a self-consistent EMC and plausible features in the transmission spectrum.  The measured strength of this possible absorption would very tentatively predict no temperature inversion, as the extra opacity source(s) needed to produce temperature inversions may tend to suppress the observability of atmospheric features, e.g., the suppressed \ion{Na}{1} absorption in HD 209458b \citep{Charbonneau2002, Snellen2008} compared to HD 189733b (R2008; Huitson et al., 2011, in preparation).

The lack of H$\alpha$ detections in these two targets as well as lack of comparison points in their atmospheric spectra in the literature of these planets severely limits our ability to draw firm conclusions about their atmospheres.  In \S\ref{ss:HD189733b}, we discuss limits on the excitation temperature, $T_{exc}$, of hydrogen due to the now-available combination of Ly$\alpha$ and H$\alpha$ measurements for HD 189733b and HD 209458b.  With Ly$\alpha$ detections, we could in principle put similar constraints on $T_{exc}$ in these two targets.

We also note here that the $a$ $priori$ expectation of detectable atmospheric absorption in HD 147506b is by far the lowest, and also lower for HD 149026b than the remaining two targets.  This is primarily based on S/N limitations due to the fact that both central stars are fainter than HD 189733 and HD 209458, which have similar apparent magnitudes---HD 149026 is fainter by $\sim$0.5$\magnitude$ and HD 147506 by $\sim$1$\magnitude$.  In addition, HD 147506b also has by far the smallest scale height of our four targets due to its very high mass even though its radius is comparable to the other planets.  In terms of scale height, HD 149026b is comparable to HD 189733b, but the prospect of detecting absorption is made more difficult due to the lower S/N resulting from the respective apparent brightness of their central stars.  Despite these relative limitations, these systems have bright enough central stars that if Ly$\alpha$ absorption exists in their exospheres, it should be detectable with a reasonable amount of observing time.  Furthermore, we note that our upper limits on H$\alpha$ are a factor of a few smaller with respect to the HD 189733b detection and the HD 209458b feature.  HD 147506b's scale height is much smaller but its Roche lobe is much larger than the other targets; this reduces the possibility that a true, unbound exosphere exists.  Our non-detection of H$\alpha$ does not confirm this, but is certainly consistent with it, and therefore not unexpected.  HD 149026b is different in that its scale height and Roche lobe are comparable to HD 189733b (slightly larger in both cases).  Therefore, our non-detection of H$\alpha$ in this target is arguably indicative of a more interesting difference between this target and HD 189733b and HD 209458b.  Whatever processes that create the features seen in these two targets (see \S\ref{ss:HD189733b} and \S\ref{ss:HD209458b}) are apparently not present, or significantly weaker, in HD 149026b.

\subsection{The HD 189733b H$\alpha$ Detection}
\label{ss:HD189733b}

\subsubsection{Identification and Description}
\label{sss:189description}
We argued in \S\ref{ss:189results} that the feature we observe in HD 189733b is highly correlated with planetary transit and is a real feature of the transmission spectrum.  If this premise is correct, two basic possibilities exist:  that the feature is the result of the influence on the stellar H$\alpha$ line from SPI, or that it is the result of absorption from the planet's atmosphere or exosphere.  HD 189733b is the most chromospherically active of the stars in our sample \citep[see][and also the discussion in \S\ref{sss:189interpretation}]{Knutson2010}, and thus the first possibility should not be ignored---especially given that we strongly consider SPI as a possible explanation for the unusual feature in HD 209458b in \S\ref{ss:HD209458b}.  However, confirmed instances of SPI are quite rare \citep[e.g.,][and references therein]{Poppenhaeger2011}.  Furthermore, in cases where SPI has been detected, the effect on the star is complex as a function of time.  For example, HD 179949, a non-transiting system, shows the influence of the planet for nearly half of its orbital cycle, and also may show additional variability on larger time scales such as the star's activity cycle \citep{Shkolnik2008}.  Recently, the possibly of HD 189733's X-ray activity being correlated with the planetary phase was suggested by \citet{Pillitteri2011}, who acknowledge that further observations are needed to confirm or reject this hypothesis.

In the simplest case, though, we might expect that the planet influences additional activity on the portion of the stellar surface that is facing the planet.  If this were the case in HD 189733, the in-transit observation would have additional chromospheric emission during transit that ``fills in" the H$\alpha$ absorption core.  This would result in an emission feature in the transmission spectrum rather than the absorption feature we see.  Therefore, the interpretation that the absorption comes from some portion of the planet's atmosphere is highly preferred.

As discussed in \S\ref{s:intro}, there are many known detections of exoplanetary atmospheres, and this includes exospheric absorption from the Ly$\alpha$ transition of hydrogen in HD 189733b and HD 209458b and the Balmer edge of \ion{H}{1} in HD 209458b.  The detection of H$\alpha$ in HD 189733b is the first such detection of this line in an exoplanetary atmosphere or exosphere.  It is also more generally the first detection of atomic material in an exoplanetary atmosphere that is in an excited state via a resolved line, and combined with the \citet{LecavelierDesEtangs2010} detection of Ly$\alpha$, results in the first case where a set of multiple, resolved lines (which do not come from the same multiplet or an ambiguous absorption edge) have been observed in a single exoplanetary atmosphere.

We detect significant H$\alpha$ absorption in HD 189733b over a range of $-24.1$ to $26.6\kmpers$.  This is determined by the continuous range over which $S_T<-3\sigma$, with the $\sigma$ here being the standard deviation of the nearby spectrum.  We note that this is much narrower than the velocity range of $-130$ to $100\kmpers$ over which \citet{VidalMadjar2003} detected Ly$\alpha$ in HD 209458b.  The \citet{LecavelierDesEtangs2010} detection of Ly$\alpha$ in HD 189733b used ACS at a resolution of $R\sim100$.  Therefore, the Ly$\alpha$ line was not resolved, and we are unable to ascertain how our detection compares to Ly$\alpha$ in this target (though we discuss this more below).  However, the contrast with the \citet{VidalMadjar2003} detection is still intriguing, because it implies at least one of the following:  (1) the Ly$\alpha$ absorption of these two targets have very different velocity ranges, or that (2) Ly$\alpha$ and H$\alpha$ absorption in HD 189733b are not over the same velocity range.

\subsubsection{Velocity Range}
\label{sss:189velocity}
The narrow velocity range over which we detect this absorption is also interesting because, at least superficially, it raises the question of whether or not we are detecting hydrogen in the exosphere.  With $R_{\rm HD \; 189733b}=1.14R_{\rm Jupiter}$ and $M_{\rm HD \; 189733b}=1.138M_{\rm Jupiter}$, the escape velocity,
\begin{equation}\label{eq:escvel}
v_{esc}=\sqrt{\frac{2GM}{R}} \; ,
\end{equation}
is 59.5$\kmpers$ (at the planetary surface).  With the \citet{LecavelierDesEtangs2010} detection of Ly$\alpha$, the absorption was not resolved over the 150$\kmpers$ full-width at half-maximum (FWHM) of the Ly$\alpha$ line in the ACS spectrum.\footnote{The instrumental resolution is $\sim$300$\kmpers$ (FWHM), but \citet{LecavelierDesEtangs2010} estimate the stellar Ly$\alpha$ width from the empirical relationship of \citet{LandsmanSimon1993}.}  \citet{LecavelierDesEtangs2010} argue that under any scenario, the hydrogen must be escaping---though unresolved, the total amount of absorption demands that either some of the gas is at a velocity greater than the escape velocity, or some of it is beyond the planet's Roche lobe.

Our detection of H$\alpha$ is not as clear in this respect.  The maximum absolute velocity at which we detect clearly H$\alpha$ absorption, $\sim$25$\kmpers$, is significantly smaller than the escape velocity at the planetary surface of $\sim$59.5$\kmpers$.  The escape velocity is equivalent to 25$\kmpers$ at approximately $5.7R_{\rm HD \; 189733b}$, or $6.4R_{\rm Jupiter}$ (which would also be outside HD 189733b's Roche lobe).  The maximum apparent planetary radius implied by the H$\alpha$ absorption is $1.35R_{\rm HD \; 189733b}$\footnote{This assumes no calibration as we calculated at the end of \S\ref{ss:189results}.  Assuming this calibration value is the same for the maximum absorption point as it is for the integrated measurement, the maximum apparent planetary radius is $1.29R_{\rm HD \; 189733b}$.}, which is not outside the Roche lobe.  Note, though, that calculating this apparent planetary radius assumes optically thick absorption; if the H$\alpha$ absorption is optically thin, it could be physically present at much greater radii.  Assuming the hydrogen responsible for the H$\alpha$ absorption has the same physical extent as the hydrogen responsible for the Ly$\alpha$ absorption, our result argues for the former case of lower-velocity hydrogen that overfills the Roche lobe rather than gas within the Roche lobe that exists at speeds greater than the escape velocity.

Due to the low resolution of the \citet{LecavelierDesEtangs2010} ACS spectra, we cannot directly compare the velocity ranges of the Ly$\alpha$ and H$\alpha$ absorption in HD 189733b.  However, we note that the interstellar Ly$\alpha$ absorption is probably saturated over the velocity range of our H$\alpha$ detection \citep[see Fig.~10 and the corresponding discussion in][]{LecavelierDesEtangs2010}.  Therefore, in order for there to be any detectable Ly$\alpha$ absorption in HD 189733b, it must indeed cover a wider velocity range than our H$\alpha$ detection, and any attempt to understand how hydrogen in the $n=2$ state is formed and sustained in the HD 189733b exosphere must recognize that Ly$\alpha$ is absorbed by hydrogen over a larger velocity range.  This also complicates the calculation of column density and excitation temperature that we perform below in \S\ref{sss:189column} and \S\ref{sss:189Texc}.

\subsubsection{Column Density}
\label{sss:189column}
If we assume that the H$\alpha$ absorption is optically thin, we can calculate an estimated column density from our absorption measurements by the equation 
\begin{equation}\label{eq:columndensity}
\eqw \left(\frac{\Delta A}{A}\right)=\frac{\pi e^2}{mc^2}f_{23}\lambda_{H\alpha}^2 N_2 \; ,
\end{equation}
where $(\Delta A/A)$ is the fractional area of the stellar disk covered by the absorbing material, $f_{23}$ is the oscillator strength of the H$\alpha$ transition, and $N_2$ is the column density of hydrogen in the $n=2$ state.  $\eqw$ is the equivalent width of the absorption, defined as
\begin{equation}\label{eq:eqw}
\eqw = \int{ \left [ \frac{F_{out}-F_{in}}{F_{out}} \right ] \; d\lambda} \; .
\end{equation}
However, we earlier defined the ``transmission spectrum," $S_T$, in Eq.~\ref{eq:st}, such that we can substitute and get
\begin{equation}\label{eq:eqwst}
\eqw = \int{-S_T \; d\lambda} \; ,
\end{equation}
which for approximately equally spaced points in $S_T$, and substituting Eq.~\ref{eq:intsig}, means that
\begin{equation}\label{eq:eqwts}
\eqw \approx -M_{abs}\Delta\lambda \; ,
\end{equation}
Note that the substitution from Eq.~\ref{eq:intsig} is not exact in that $M_{abs}$ only approximates $<\! \!S_T\! \!>$, because the former also includes normalization from the comparison bands.  In Table \ref{table:189results}, we show the values of $M_{abs}$ for HD 189733b in three additional integration bands beyond those shown in Table \ref{table:results}.  Table \ref{table:189results} also shows the resulting $\eqw$ of these three values of $M_{abs}$.  As should be the case, the $\eqw$ is constant to within the errors, because the additional integration region of the wider bands is only adding, in principle, $<\! \!S_T\! \!> \approx 0$.

To calculate $N_2$ is fairly straightforward with the exception of calculating $(\Delta A/A)$, because, as we have previously noted, the Ly$\alpha$ absorption---which would be our best indicator of the extent of the hydrogen envelope---is not resolved.  Using an equivalent width of 0.0128 \AA{} from the top row of Table \ref{table:189results} implies a column density of $6.4\times10^{10} {\rm \; cm}^2 \; (\Delta A/A)$.  \citet{LecavelierDesEtangs2010} quote values for the absorption, which is equivalent to $(\Delta A/A)$, as small as 5\% over the entire unresolved Ly$\alpha$ line, and as large as 30\% if restricted to the velocity range of $\pm 49\kmpers$ (the escape velocity at 1.65$R_{\rm Jupiter}$, the minimum size of the hydrogen cloud allowable by the absorption).  Thus our effective $n=2$ hydrogen column density, $N_2$, could be anywhere from $3.2\times10^{9} {\rm \; cm}^2$ to $1.9\times10^{10} {\rm \; cm}^2$.  Notably, there are still strong assumptions in these values.  The 30\% figure could be even larger if the velocity range of Ly$\alpha$ is even more restricted than $\pm 49\kmpers$, meaning these calculations are upper limits.  We have also argued that H$\alpha$ absorption occurs over a smaller velocity range than Ly$\alpha$, and therefore may not share its physical extent.  This means that $(\Delta A/A)_{\rm H\alpha} \leq (\Delta A/A)_{\rm Ly\alpha}$, and the prior calculations are instead lower limits for a given $(\Delta A/A)_{\rm Ly\alpha}$.

\subsubsection{Excitation Temperature}
\label{sss:189Texc}
The excitation, or population, temperature of an atomic or molecular species is defined as
\begin{equation}\label{eq:Texc1}
\frac{N_2}{N_1} = \frac{g_2}{g_1}\exp{\left[-\frac{\Delta E_{21}}{k_{B}T_{exc}}\right]} \; .
\end{equation}
Solving for $T_{exc}$ and plugging in constants we obtain the relationship
\begin{equation}\label{eq:Texc2}
T_{exc} = \frac{1.18\times10^5 \rm \; K}{\ln{\left[4\frac{N_1}{N_2}\right]}} \; .
\end{equation}
We have already seen above how many assumptions are in the calculation of $N_2$, but we can simplify the calculation in our approximation for $N_1$.  The resolution of the ACS in spectroscopic mode as used by \citet{LecavelierDesEtangs2010} is $R\sim100$ (FWHM), or $\sim$12 \AA{}.  They detect 5.05\% absorption over this line, but if normalized to the baseline transit depth of 2.4\%, the differential absorption is only 2.65\%.  Therefore the equivalent width is 0.32 \AA{}.  If the $b$-value\footnote{The term ``$b$-value" refers to the Doppler parameter, $b=\sigma\sqrt{2}$, of a Gaussian velocity distribution.} of the exospheric hydrogen is large enough that the absorption is not saturated then we can approximate $N_1/N_2$ with $W_{\lambda, \; {\rm Ly\alpha}}/W_{\lambda, \; {\rm H\alpha}}$.  This results in a temperature of $2.6\times10^4$ K.  Of course, we are again violating our previous conclusion that Ly$\alpha$ and H$\alpha$ are not fully spatially coincident.  We can explicitly state this assumption and write the excitation temperature as
\begin{equation}\label{eq:Texc3}
T_{exc} = \frac{1.18\times10^5 \rm \; K}{4.61+\ln{\left[\frac{\Delta A_{\rm Ly\alpha}}{\Delta A_{\rm H\alpha}}\right]}} \; .
\end{equation}
In this case, even a spatial discrepancy of a factor of 3 between Ly$\alpha$ and H$\alpha$ changes the excitation temperature by $<$25\%.

\begin{figure}[t!]
\begin{center}
\epsscale{1.00}
\plotone{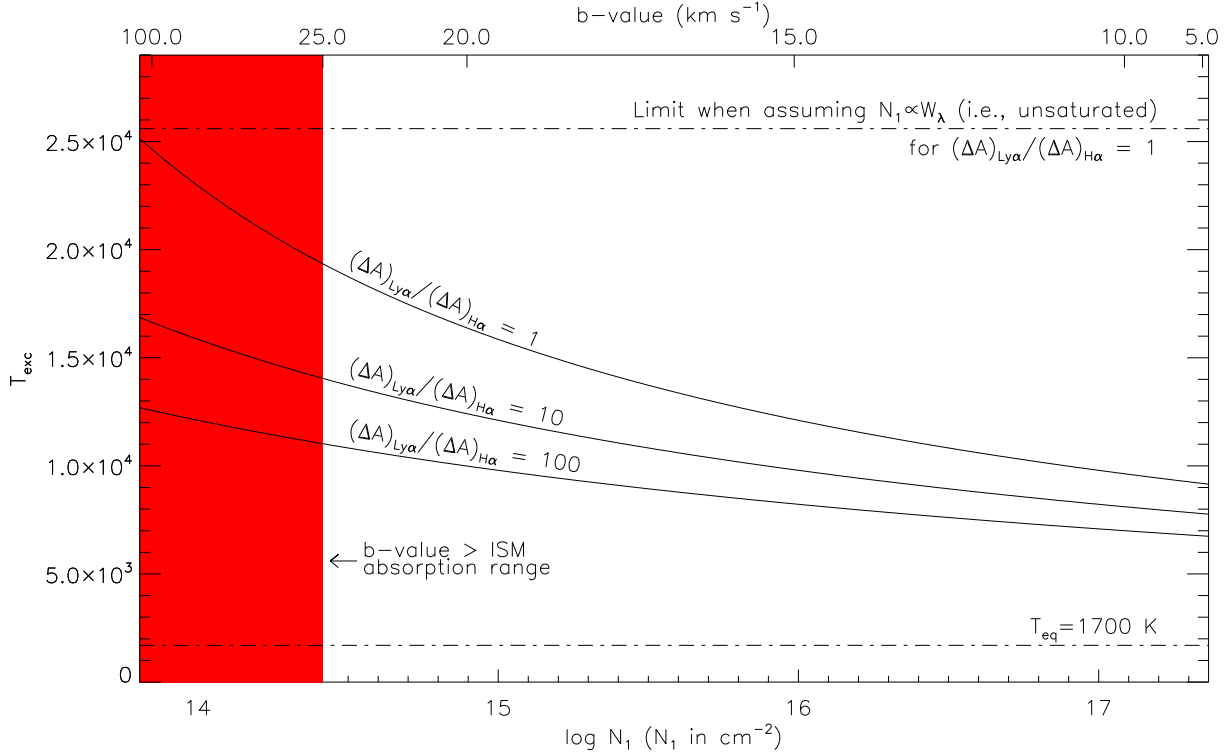}
\end{center}
\caption[]{A plot showing how our derived $T_{exc}$ varies as a function of $n=1$ column density $N_1$ (based on the Ly$\alpha$ absorption measurements), as well as $(\Delta A)_{\rm Ly\alpha}/(\Delta A)_{\rm H\alpha}$.  The top axis shows the $b$-values (i.e., the Doppler parameter $b=\sigma\sqrt{2}$ for a Gaussian velocity distribution) that correspond to these column densities.  As $b \rightarrow \infty$, the absorption is unsaturated and a lower limit on $N_1$ is reached; the left-hand axis is at $b=200\kmpers$ and nearly at this limit.  Also shown are $T_{eq}$ (defined in \S\ref{sss:189interpretation}) and the limiting $T_{exc}$ if the Ly$\alpha$ $\eqw$ of \citet{LecavelierDesEtangs2010} is an unsaturated measurement (which is unlikely).  The red shaded area is the range of $b$-values roughly implied by the fact that in Fig.~10 of \citet{LecavelierDesEtangs2010} the ISM absorption is assumed to be saturated over the range $\pm25\kmpers$.  However, this is not a robust inference as a damped Ly$\alpha$ line could show absorption at higher velocities even at low $b$.  Note that if the H$\alpha$ absorption is saturated, $T_{exc}$ increases; if Ly$\alpha$ absorption is underestimated due to the limits of the low-resolution ACS data (which is almost certain), then $T_{exc}$ decreases.}
\label{fig:texc}
\end{figure}

It is very likely that the Ly$\alpha$ absorption is indeed saturated, which would make $N_1$ larger and therefore our calculation of $T_{exc}$ is an upper limit.  To quantify this would require another additive term of $\ln{\left( N_{\rm 1, \; actual}/N_{\rm 1, \; unsaturated} \right) }$ in the denominator of Eq.~\ref{eq:Texc3}.  Fig.~\ref{fig:texc} shows the dependence of $T_{exc}$ on $N_1$ (and implicitly on $b$-value) as well as $\Delta A_{\rm Ly\alpha} / \Delta A_{\rm H\alpha}$.  However, there are two important fixed assumptions in Fig.~\ref{fig:texc}:  one is that the Ly$\alpha$ $\eqw$ is not corrected for ISM absorption, and the other is that the H$\alpha$ absorption is unsaturated.  Both assumptions mean that the $N_1$ and $N_2$ column densities are lower limits, which would respectively decrease or increase the calculated $T_{exc}$.  Notably, \citet{Ballester2007} argue that any H$\alpha$ absorption in HD 209458b would be optically thick and in a thin layer (or $\Delta A_{\rm Ly\alpha} / \Delta A_{\rm H\alpha} \gg 1$ in our discussion here); if this is true of HD 189733b, these two facts would have competing effects on $T_{exc}$ relative to our assumptions.


\subsubsection{Physical Interpretation}
\label{sss:189interpretation}
What sort of physical model can explain these conditions?  If we assume that HD 189733b has zero albedo and the surface is in radiative equilibrium with the star as its only energy source, we can calculate a temperature of 1700 K (which we will refer to as $T_{eq}$).  This is an order of magnitude smaller than the excitation temperature that we have calculated.  The uncertain quantities that would reduce our final calculation of $T_{exc}$ are $N_1$, $(\Delta A/A)_{\rm H\alpha}$, and $(\Delta A/A)_{\rm Ly\alpha}$; $T_{exc}$ depends on the logarithm of these numbers.  Therefore, we would need to be off by a total of $\sim$28 orders of magnitude in these quantities in order to make $T_{exc}=T_{eq}=1700$ K.  If we have underestimated the effect of these quantities by a total of only five orders of magnitude, $T_{exc}$ is still greater than $T_{eq}$ by a factor of more than four.

The discrepancy between $T_{eq}$ and $T_{exc}$ argues first that the hydrogen that produces our observed H$\alpha$ absorption must be in an extended, low density portion of the atmosphere.  If it were too close to the surface, it would likely be in thermal equilibrium with the planet at a value close to $T_{eq}$; if it were at higher density, it would also likely be collisionally de-excited.  Therefore, the likely source of excited $n=2$ hydrogen is through radiation.  Of our four systems, HD 189733 is the latest spectral type, an early K star (HD 149026 is a G0 star, while HD 147506 and HD 209458 are F8 stars).  However, it also has the smallest planet-star separation, and has the most active \ion{Ca}{2} H \& K emission.  HD 189733's measured value of $S_{\rm HK}$---a quantity indicating stellar activity---is 0.508, compared to 0.191, 0.152, and 0.160 in HD 147506, HD 149026, and HD 209458, respectively \citep{Knutson2010}.  Similarly, HD 189733's value of $\log{\left( R'_{\rm HK} \right)}$---a quantity derived from $S_{\rm HK}$ that allows for better comparison between different spectral types---is $-4.501$, compared to $-4.780$, $-5.030$, and $-4.970$.  \citet{Knutson2010} also note that HD 147506, with the second highest value of $\log{\left( R'_{\rm HK} \right)}$ in our target list, is at the edge of the $B-V$ range over which the calibration between $S_{\rm HK}$ and $\log{\left( R'_{\rm HK} \right)}$ is valid.  They subsequently argue, based on its spectra at \ion{Ca}{2} H \& K, that HD 147506 has a quieter chromosphere than its derived $\log{\left( R'_{\rm HK} \right)}$ value would otherwise suggest.

The importance of \ion{Ca}{2} H \& K emission is that it is strongly correlated with general chromospheric activity, and we would expect it to therefore correlate strongly with Ly$\alpha$ and UV activity, although the directly measured evidence for this is limited by a lack of surveys in the literature.  Stellar Ly$\alpha$ and UV activity would be able to populate the $n=2$ state of hydrogen.  Ly$\alpha$ emission would directly create $n=2$ hydrogen in the planet's atmosphere, while higher-frequency UV emission would populate the $n>2$ states, which would then radiatively cascade down to $n=2$.

If we assume that radiation is the dominant excitation mechanism, this means that the outermost and star-facing portion of HD 189733b's atmosphere will be exposed to the most radiation.  This portion of the atmosphere will also be at the lowest density.  Therefore, it is possible for some hydrogen to remain in an excited states at high altitudes, whereas at lower altitudes it will perhaps collisionally de-excite.  If the $n=2$ hydrogen is only at the highest altitudes of HD 189733b's atmosphere, this may also explain its comparatively restricted velocity range---if it is already beyond the planet's Roche lobe (as we argued in \S\ref{sss:189velocity}), it is not bound to the planet, and there is no physical or observational requirements that it be at a high velocity.  The low density of this outer layer also means that it may be at a lower pressure than the rest of the atmosphere, consistent with the smaller velocity range.

The best way to better understand the H$\alpha$ absorption in HD 189733b's exosphere would be to explore its Ly$\alpha$ absorption in more detail.  \citet{VidalMadjar2003} were able to study HD 209458b's Ly$\alpha$ absorption in great detail due to their use of the E140M grating of STIS, whereas \citet{LecavelierDesEtangs2010} were unable to resolve the Ly$\alpha$ absorption (using ACS while STIS was not operational).  Observing HD 189733b's Ly$\alpha$ absorption at a resolution comparable to \citet{VidalMadjar2003} would enable a direct comparison of the velocity range over which each is present, and allow for a more detailed model of the Ly$\alpha$ absorption (incorporating interstellar absorption, for example) and the resulting column density.

\subsection{The HD 209458b Feature}
\label{ss:HD209458b}

\subsubsection{Identification and Description}
\label{sss:209description}
In \S\ref{ss:209results} we noted the unusual, strong feature in the transmission spectrum of HD 209458b.  We noted above that this feature is correlated with transit (or at least planetary phase) in our EMC results.  All the reduction steps are performed ``blindly" with no knowledge of whether an individual observation is in- or out-of-transit prior to the coaddition step; furthermore, we note that none of the other targets showed a similar feature in their master transit spectra.  This argues that this feature is a real, planetary phase-correlated feature of unknown origin.  However, we cannot completely rule out some unknown reduction effect, perhaps that is uniquely affected by some characteristic of the HD 209458b observations.

We previously mentioned two broad categories of solutions that could be responsible for this feature---the Rossiter-McLaughlin Effect (RME), which creates a variation in the transmission spectrum that is dependent on wavelength, or a misalignment of the broad component of the H$\alpha$ profile between the in- and out-of-transit spectra that is relative to the H$\alpha$ line core and therefore not addressed by our process of higher-order wavelength correction.  While these processes have the potential to match qualitatively what we see in HD 209458b, we did not assess the quantitative plausibility of either scenario, which we do here.

\subsubsection{Rossiter-McLaughlin Effect?}
\label{sss:209rmeffect}
The RME is able to create features in a transmission spectrum that qualitatively match what we see---dips and/or spikes near stellar absorption lines that are the result of the planet blocking different portions of the star's spectrum in velocity space.\footnote{Note that this is superficially similar and related to, but nevertheless fundamentally different from, the feature that results from the RME in the more common plot of radial velocity as a function of time.}  To test this, we made a simple simulation of the RME, in which we made a simulated H$\alpha$ profile, assumed the stellar rotation speed, and calculated the RME based on the distribution of our observations during transit (i.e., which part of the velocity profile of the star is blocked in each observation).  Under a wide range of reasonable assumptions for this model (e.g.,~limb darkening parameters), we find multiple bases on which to reject the hypothesis that this large feature is the direct result of the RME.  The first argument against this scenario is that the in-transit observations of HD 209458 are (randomly) weighted toward the first half of transit; when we simulate the RME, this results in spectral features with the opposite blue/red characteristics as the feature in the HD 209458b transmission spectrum.  That is to say, features in the first half of transit block the blueshifted portion of the stellar disk, and should result in a ``dip" to the blue and a ``spike" to the red---the opposite of what we see.  Furthermore, the feature we see in the HD 209458b transmission spectrum is too broad, by perhaps an order or magnitude or more, to be caused by the RME in a system where the star has a small known $v \sin{i}$ of only a few ${\rm km \; s^{-1}}$.  Furthermore, in our simulations, the vertical magnitude of the feature (i.e., the height of the spike and dip) is also much smaller than the observed feature, again by a factor of several (the simulations would indicate that the magnitude of any RME is within the noise shown in Fig.~\ref{fig:hadiffspec}).  Most importantly, we note that the RME is purely a transit effect, and we discussed in \S\ref{ss:209results} and showed in Fig.~\ref{fig:209phase} that the effect may be a more general phase effect with a maximum near transit.  Thus we can rule out the RME as being responsible for this feature.

\subsubsection{Broad Stellar H$\alpha$ Component Shifting?}
\label{sss:209broadhashift}
An alternative scenario for this feature is a relative velocity shift between the narrow and broad components of the stellar H$\alpha$ line that is correlated with planetary phase.  We made a simple model of the H$\alpha$ line by modeling it as two Gaussian components, one narrow and one broad.  Using this method we fitted the master in-transit and out-of-transit spectra.  When the narrow components of the master in- and out-of-transit spectra are aligned, the broad component of the out-of-transit spectrum is shifted by approximately $-0.08$ \AA{} (or $-4\kmpers$) relative to the in-transit spectrum.  This model is a qualitative visual match to the spectrum, though the fits are statistically poor ($\chi_{\nu}^2 \gg 1$) due to the crudeness of this toy model.  However, when we take the models obtained from these simple double-Gaussian model fits, one for the master in-transit spectrum and one for the master out-of-transit spectrum, and use them to create a synthetic transmission spectrum, we are able to replicate the observed transmission spectrum with $\chi_{\nu}^2 < 2$.  Thus it seems reasonably clear that this model can produce this feature in a way that, as described above in \S\ref{sss:209rmeffect}, the RME cannot.

We note that the $-4\kmpers$ shift is on the same order as the instrumental resolution of  $5\kmpers$, though we are essentially performing a centroid measurement, which can be done with greater precision than the FWHM of an instrument's line spread function.  This again raises the issue of a reduction artifact, which we discussed in \S\ref{ss:209results} and \S\ref{sss:209description}, but here we explore real, physical processes that might be responsible for such a shift that is strongly correlated with planetary transit.

Some sort of SPI could conceivably cause this shift.  As pointed out earlier, the evidence for SPI is very limited.  We do note that highly interactive RS CVn binary systems display line behavior that is qualitatively similar to the feature in the transmission spectrum of HD 209458b \citep[e.g., the observed vs.~theoretical residuals of H$\alpha$ in IM Peg in][]{Biazzo2006}.  Another extremely similar feature is shown in \citet{HallRamsey1992} in the ``subtracted spectrum" shown for an eclipse of the AW Her system; this ``subtracted spectrum" is the ratio of H$\alpha$ to H$\beta$.  Neither of these cases are perfect physical analogues to a transit and our transmission spectrum---particularly the \citet{HallRamsey1992} result, which is a ratio of two lines rather than a comparison of the same line in- vs.~out-of-transit---but they do show characteristics that are similar to what we observe.

The two primary ways in which a planet could influence its central star are through tidal and magnetic interactions.  A tidal interaction could produce tidal bulges on the star, resulting in increased stellar activity due to photospheric turbulence \citep{Cuntz2000, Poppenhaeger2011}.  Magnetic interactions can increase stellar activity through increased magnetic reconnection \citep{Lanza2008} or heating of the stellar atmosphere by flux tubes analogous to those in the Jupiter-Io system \citep{Schmitt2009}.  If there is induced stellar activity on the surface due to the planet, it may produce emission that alters the shape of the broad or narrow component of the H$\alpha$ as a function of whether the system is in a state of in- vs.~out-of-transit.  For example, residual \ion{Ca}{2} flux in HD 189733 \citep{Shkolnik2008} shows shifts in wavelength of up to $\sim$0.25 \AA{} (see Fig.~4 of that paper).  Perhaps an analogous residual flux in HD 209458's H$\alpha$ line is of nearly constant intensity when averaged over the many observations but has a slightly shifted wavelength when the in-transit observations are compared to the out-of-transit observations.  This shifting flux, imposed on the stellar line, might be able to create the narrow/broad relative shift that we see.  Another possibility is that the magnetic influence of the planet directly influences the broad component of the stellar H$\alpha$ line.  If there is any physical separation in the layers of the stellar line---e.g., the core is dominated by a different layer than the wings---then this may be plausible.

One difficulty with invoking stellar activity is that HD 209458 is not known to be particularly active, and is much less active than HD 189733.  Why HD 209458, a less active star with a more distant planet (thus reducing the likelihood of SPI), would show this sort of effect is unclear.  

\subsubsection{Shifting Due to an Existing Hydrogen Torus?}
\label{sss:209Hatorus}
A related alternative to the idea that SPI is causing the shifting of broad-component H$\alpha$ in the star is the possibility that a component of the apparent stellar H$\alpha$ line has a planetary origin.  In other words, perhaps there a collection of material around the star, in a roughly toroidal shape, near the planet's orbit that formed from the escaping material from the planet.  This would require that some of this hydrogen is excited and capable of exhibiting H$\alpha$ absorption and that the planetary orbit affects the torus in some way related to its orbit.  If this is a part of the broad component of the apparent stellar H$\alpha$ profile, and it shifts in response to the planetary orbit, then it could be responsible for the feature that we see in the transmission spectrum.

One difficulty with this interpretation is the extent of the torus.  If it is only a partial torus near the planet, or a ``comet-like" shape of material trailing the planet, then we would still expect to see an absorption feature in the master transmission spectrum.  Given that we do not see such an absorption feature, for this general scenario of a torus of material to be correct would imply that the material exists continuously throughout the planetary orbit, and experiences a velocity shift in response to the planetary orbit.  This would pose the problem of how the $n=2$ hydrogen is created and sustained in the excited state.  It would also be unclear why the shift would be similar at transit and at secondary eclipse, and different at quadrature.  Further observations are needed to confirm or reject the feature that we see in HD 209458b, and, if confirmed, determine its nature and origin.


\section{SUMMARY}
\label{s:summary}
We perform atmospheric transmission spectroscopy of four exoplanetary targets in the spectral region surrounding the H$\alpha$ line.  We find a strong feature of apparent H$\alpha$ absorption in the transmission spectrum of HD 189733b.  Our best attempts to remove stellar variability and assess other systematic errors indicate that this feature is real and correlated with planetary transit.  The level of absorption is $(-8.72\pm1.48)\times10^{-4}$ integrated over a 16 \AA{} band and relative to the adjacent spectrum, significant at a 5.9$\sigma$ level.  Several complications exist in using this absorption to derive $T_{exc}$.  We estimate $T_{exc}=2.6\times10^4$ after employing some simplifying assumptions, but discuss these assumptions and their effects in detail.  In HD 209458b we find a dramatic feature with reflectional symmetry and discuss possible explanations.  While we are unable to give a clear physical explanation for this feature, we discuss the clear correlation with planetary orbital phase.  We place upper limits on the H$\alpha$ absorption in our two remaining targets.  These results all lend themselves to follow-up work to confirm and/or refine the properties of the observed features in HD 189733b and HD 209458b, or to search for hydrogen absorption in other transitions of all four of these targets in order to better understand the nature hydrogen excitation in exoplanetary atmospheres, as well as possible insights into atmospheric escape in these planets.

\acknowledgements
We thank the anonymous referee, Jeffrey Linsky, and Thomas Ayres for providing several helpful comments on the manuscript.  Authors A.~G.~J.~and S.~R.~acknowledge support by the National Science Foundation through Astronomy and Astrophysics Research Grant AST-0903573 (PI:  S.~R.).  The Hobby-Eberly Telescope is a joint project of the University of Texas at Austin, the Pennsylvania State University, Stanford University, Ludwig-Maximilians-Universit\"{a}t M\"{u}nchen, and Georg-August-Universit\"{a}t G\"{o}ttingen and is named in honor of its principal benefactors, William P.~Hobby and Robert E.~Eberly.  This work made use of IDL, the Interactive Data Language;\footnote{http://www.ittvis.com/language/en-us/productsservices/idl.aspx} IRAF, the Image Reduction and Analysis Facility;\footnote{http://iraf.noao.edu/} the SIMBAD Database;\footnote{http://simbad.u-strasbg.fr/simbad/} the Exoplanet Data Explorer \citep{Wright2011};\footnote{http://exoplanets.org/} and the Exoplanet Transit Database.\footnote{http://var2.astro.cz/ETD/index.php}

\bibliographystyle{apj}
\bibliography{refs}

\clearpage \clearpage

\clearpage \clearpage

\clearpage \clearpage

\clearpage \clearpage

\clearpage \clearpage

\clearpage \clearpage

\clearpage \clearpage

\clearpage \clearpage

\clearpage \clearpage

\clearpage \clearpage

\begin{deluxetable}{cccccccccc}
\tablecolumns{10}
\tablewidth{0pc}
\rotate
\tabletypesize{\scriptsize}
\tablecaption{Observations\label{table:observations}}
\tablehead{\colhead{System} & \multicolumn{2}{c}{In Transit\tablenotemark{a}} & \multicolumn{2}{c}{Out of Transit\tablenotemark{a}} & \multicolumn{2}{c}{Ingress/Egress\tablenotemark{a}} & \colhead{Telluric} & \multicolumn{2}{c}{Observations} \\
\colhead{} & \colhead {\# of Obs.} & \colhead{Time (s)\tablenotemark{b}} & \colhead {\# of Obs.} & \colhead{Time (s)\tablenotemark{b}} & \colhead{\# of Obs.} & \colhead{Time (s)\tablenotemark{b}} & \colhead{} & \colhead{\# of Obs.} & \colhead{$<$Time$>$ (s)\tablenotemark{b}}}
\startdata
HD 147506\tablenotemark{c} & 28 & 16508 & 84 & 49687 & 5 & 3000 & HD 156729\tablenotemark{d} & 34 & 567 \\
HD 149026 & 22 (1) & 12718 (600) & 90 & 52734 & 5 & 3000 & HD 156729\tablenotemark{d} & 34 & 567 \\
HD 189733\tablenotemark{e} & 38 (3) & 20140 (1640) & 173 (1) & 79011 (0) & 15 (1) & 8243 (600) & HD 196867 & 45 & 291 \\
HD 209458 & 29 (6) & 17310 (3510) & 72 (3) & 43175 (1800) & 4 & 2400 & HD 218045 & 21 & 80 \\
\enddata
\tablenotetext{a}{Numbers shown are all observations taken, while the numbers in parentheses indicate observations that are discarded for various reasons, primary cloud-induced low S/N, but also various reduction artifacts.  The ``In Transit" and ``Out of Transit" columns are mutually exclusive, defined by observation midpoint, while the ``Ingress/Egress" observations are a subset of the ``In Transit" observations, again defined by observation midpoint.}
\tablenotetext{b}{Times for primary science target exposures are total times, while telluric exposure times are average times, as the telluric observations are always used independently and are never coadded.}
\tablenotetext{c}{HD 147506 and HD 147506b are also known as HAT-P-2 and HAT-P-2b, respectively.}
\tablenotetext{d}{Observations of HD 147506 and HD 149026 use the same telluric star.  The values in the table this telluric represent the same observations, i.e.~the 34 observations listed do not distinguish between which primary science target was originally associated with the telluric observation.}
\tablenotetext{e}{Some IRAF headers from the queue observing program prior to 12/2007 have ``0 s" incorrectly listed in the exposure time header.  This is true for 10 out of 211 observations of the HD 189733 system, and no observations for the other targets.  Therefore, we estimate that the times shown for HD 189733 are incomplete by as much as 1800 s of in transit time (none discarded), 3000 s of out of transit time (600 s discarded, corresponding to the ``0" in the table above), and 600 s of ingress/egress time (none discarded).}
\end{deluxetable}

\begin{deluxetable}{cccccccc}
\tablecolumns{8}
\tablewidth{0pc}
\rotate
\tabletypesize{\scriptsize}
\tablecaption{System and Orbital Parameters of Targets\label{table:systemparams}}
\tablehead{\colhead{Planet} & \colhead{Reference Epoch} & \colhead{Period} & \colhead{Transit Duration} & \colhead{$a$/R$_{\rm *}$} & \colhead{R$_{\rm P}$/R$_{\rm *}$} & \colhead{Inclination} & \colhead{References}\\
\colhead{} & \colhead{(HJD$-2400000$)} & \colhead{(days)} & \colhead{(minutes)} & \colhead{} & \colhead{} & \colhead{($^{\circ}$)}}
\startdata
HD 147506b & $54387.4937\pm0.00074$      & $5.633473\pm0.0000061$     & $257.76\pm1.87$ & $8.99^{+0.39}_{-0.41}$ & $0.07227\pm0.00061$ & $86.7^{+1.12}_{-0.87}$ & 1, 2, 3, 4\\
HD 149026b & $54456.7876\pm0.00014$      & $2.87589\pm0.00001$            & $194.4\pm8.7$      & $7.11^{+0.03}_{-0.81}$  & $0.0491^{+0.0018}_{-0.0005}$ & $90.0^{+0.0}_{-3.0}$ & 3, 5, 6\\
HD 189733b & $54279.43671\pm0.000015$ & $2.2185757\pm0.00000015$ & $106.1\pm0.7$      & $8.81\pm0.06$              & $0.15436\pm0.00022$ & $85.58\pm0.06$ & 3, 7, 8, 9 \\
HD 209458b & $52854.825\pm0.000135$      & $3.5247455\pm0.00000018$ & $184.2\pm1.6$      & $8.76\pm0.04$              & $0.12086\pm0.0001$ & $86.71\pm0.05$ & 3, 10 \\
\enddata
\tablerefs{(1) \citet{Pal2010}.  (2) \citet{Bakos2010}. (3) \citet{Torres2008}.  (4) \citet{Winn2007}.  (5) \citet{Carter2009}.  (6) \citet{Wolf2007}.  (7) \citet{Agol2010}.  (8) \citet{Triaud2009}.  (9) \citet{Bouchy2005}.  (10) \citet{Winn2005}.  This table also made use of the Exoplanet Data Explorer \citep[][http://exoplanets.org]{Wright2011} and the Exoplanet Transit Database (http://var2.astro.cz/ETD/index.php).}
\end{deluxetable}

\begin{deluxetable}{crr}
\tablecolumns{3}
\tablewidth{0pc}
\tabletypesize{\small}
\tablecaption{H$\alpha$ Absorption Results\label{table:results}}
\tablehead{\colhead{Planet} & \colhead{16 \AA{} Band} & \colhead{6 \AA{} Band}}
\startdata
HD 147506b & $(-4.90\pm1.75)\times10^{-4}$ & $(-2.87\pm4.97)\times10^{-4}$ \\
HD 149026b & $(3.07\pm1.53)\times10^{-4}$ & $(7.76\pm4.20)\times10^{-4}$ \\
HD 189733b\tablenotemark{a} & $(-10.6\pm2.43)\times10^{-4}$ & $(-30.2\pm6.33)\times10^{-4}$ \\
HD 209458b & $(0.503\pm0.802)\times10^{-4}$ & $(6.91\pm2.01)\times10^{-4}$ \\
\enddata
\tablenotetext{a}{Uncalibrated values; see \S\ref{ss:189results} and Table \ref{table:189results}.}
\end{deluxetable}

\begin{deluxetable}{crrl}
\tablecolumns{4}
\tablewidth{0pc}
\tabletypesize{\scriptsize}
\tablecaption{Additional HD 189733b H$\alpha$ Absorption Results\label{table:189results}}
\tablehead{\colhead{Velocity Range (${\rm km \; s^{-1}}$)} & \colhead{$\frac{M_{abs}}{10^{-4}}$\tablenotemark{a}} & \colhead{$\eqw$ (\AA{})\tablenotemark{a}} & \colhead{Physical Significance}}
\startdata
$[-24.1,26.6]$ & $-143$ $(-118)$ & 0.0155 (0.0128) & Range of H$\alpha$ detection, this work \\ 
$[-59.5,59.5]$ & $-61.6$ $(-50.7)$ & 0.0159 (0.0131) & $\pm v_{esc}$ at HD 189733b surface \\ 
$[-150,150]$   & $-27.5$ $(-22.6)$ & 0.0180 (0.0148) & Range (FWHM) of unresolved Ly$\alpha$ emission \\ 
$[-365,365]$   & $-10.7$ $(-8.8)$ & 0.0171 (0.0141) & Initial integration band of 16 \AA{} centered on H$\alpha$ \\ 
\enddata
\tablenotetext{a}{Calibrated values in parentheses; see \S\ref{ss:189results}.}
\end{deluxetable}

\clearpage \clearpage

\end{document}